\documentclass[12pt]{article}

\usepackage{amssymb}
\usepackage{amsfonts}
\usepackage{latexsym}
\usepackage{graphicx}

\usepackage[usenames]{color}
\usepackage{amsmath,amsthm}

\usepackage{epsfig}

\usepackage{color}
\usepackage{amsmath,amsbsy,amsfonts,amssymb,latexsym,graphics,epsfig}

\newcommand{\ignore}[1]{}

\newcommand{\beqn}{\begin{eqnarray*}}
\newcommand{\eeqn}{\end{eqnarray*}}

\newcommand{\hd}{\mathcal{H}}
\newcommand{\tl}{\mathcal{T}}

\newcommand{\ee}{\mathrm{e}}

\newcommand{\Z}{\mathbb{Z}}

\newcommand{\R}{\mathbb{R}}

\newcommand{\sot}{\mbox{\footnotesize $\frac{1}{3}$}}

\newcommand{\CC}{{\mathcal C}}

\newcommand{\GG}{{\mathcal G}}

\newcommand{\KK}{{\mathcal K}}
\newcommand{\MM}{{\mathcal M}}

\newtheorem{theorem}{Theorem}[section]

\newtheorem{definition}[theorem]{Definition}

\newtheorem{remark}[theorem]{Remark}

\newtheorem{example}[theorem]{Example}

\renewcommand{\epsilon}{\varepsilon}
\newcommand{\Sone}{{\mathbb S}^1}

\newcommand{\ii}{\mbox{i}}

{\begingroup

  \begin{enumerate}}
  {\end{enumerate}\endgroup}

\newcommand{\Dr}{\mathrm{D}}

\title{Synchronous Propagation of Periodic Signals 
in Feedforward Networks of Standard Model Neurons}
\author{Ian Stewart  and David Wood\\ Mathematics Institute\\ University of Warwick \\ Coventry CV4 7AL
\\ United Kingdom}

\date{\today}

\begin{document}
\maketitle

\begin{abstract}
Periodic signals propagating along chains are common in
biology, for example in locomotion and peristalsis, and are also of interest for
continuum robots. In previous work we constructed such networks
as `feedforward lifts' of a central pattern generator (CPG). 
When the CPG undergoes periodic oscillations, created by Hopf bifurcation or
other mechanisms, it can then
transmit periodic signals along one or more feedforward chains in a
synchronous or phase-synchronous manner.
We proved necessary and sufficient conditions for
the stability of these lifted periodic orbits, in several senses. 
Here we examine the implications
of the resulting theory for chains of neurons, using several standard 
neuron models: FitzHugh--Nagumo, Morris--Lecar, Hindmarsh--Rose, 
and Hodgkin--Huxley. We
compare different notions of transverse stability, and summarize some
numerical simulations showing that for all these neuron models
the propagating signal can be transversely Floquet stable.
Finally we discuss implications for less idealized models.
\end{abstract}


\section{Introduction}
\label{S:intro}

This paper is a continuation of \cite{SW23a}, which discussed theoretical aspects
of signal propagation in a {\em feedforward lift}. This is a network in which
a small subnetwork (central pattern generator or CPG) sends signals to a longer
chain of repeating modules, constructed so that nodes along the chain can be 
synchronous with nodes of the CPG. (For simplicity we often use the term `chain'
without implying strict linear topology. The results apply to any feedforward lift,
for example a tree. Thus the propagating signal can branch.) The CPG dynamic
can be steady, oscillatory, or even chaotic. We concentrate on the periodic
case, where nodes can also be phase-synchronous ; that is, synchronous
except for a phase shift. One natural way to generate periodic
oscillations is via Hopf bifurcation but other mechanisms are also possible;
our results apply to any periodic state, however it is created.

The analysis in \cite{SW23a} is based on
the formal theory of network dynamics of \cite{SGP03,GST05,GS23}, and the
CPG is both a quotient network and a subnetwork of the full network. 
The formal theory is `model-independent' in the sense that the results do not
depend on specific model equations. However, they cannot be {\em completely} independent
of the model; certain generic
hypotheses must be satisfied, in the same way that Hopf bifurcation 
to a periodic orbit is generic if critical eigenvalues are simple, purely imaginary, 
non-resonant, and satisfy the eigenvalue-crossing condition \cite{GH83,HKW81}.

\begin{figure}[h!]
\centerline{%
\includegraphics[width=0.6\textwidth]{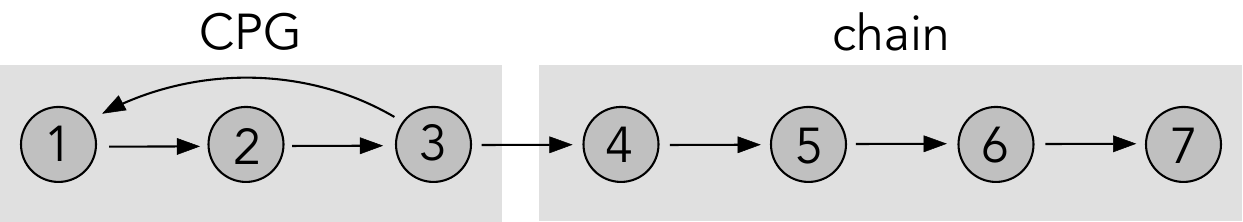}
}
\vspace{10pt}
\centerline{%
\includegraphics[width=0.6\textwidth]{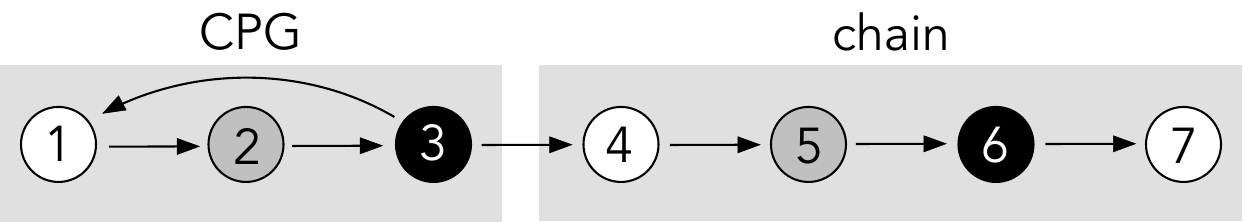}
}
\caption{A 7-node network with
one node-type and one-arrow type. {\em Top}: A balanced coloring 
corresponding to dynamics in which all nodes are synchronous. When
nodes oscillate this gives a standing wave along the chain.
{\em Bottom}: A balanced coloring 
corresponding to dynamics in which the nodes fall into
three synchrony clusters $\{1,4,7\}, \{2,5\}$, and $\{3,6\}$,
with colors white, gray, and black. 
The CPG $\{1,2,3\}$
has cyclic group symmetry $\Z_3$, so this network naturally 
propagates signals such that nodes of the same color are in synchrony 
and successive nodes are phase-shifted by $\sot$ of the period.
When nodes oscillate this appears to create a traveling wave along the chain.}
\label{F:7nodeFFZ3}
\end{figure}

Figure~\ref{F:7nodeFFZ3}
is a simple `toy' example used in \cite{SW23a} to illustrate the main ideas,
and we continue to use it here.
The CPG is the subnetwork with nodes $\{1,2,3\}$
and the arrows connecting them; the chain comprises the remainder of the network.
We emphasize that the stability computations reported below 
for this model remain valid if the chain is replaced by {\em any} feedforward lift.
However, the results depend on the chosen model for CPG dynamics.

Intuitively, nodes 1 and 4 receive the same signal from node 3 and have the same internal dynamics,
so they can synchronize. Then nodes 2 and 5 receive synchronous signals,
so they can synchronize. Similarly node 6 can synchronize with node 3, node 7
with node 4, and so on for longer chains. More formally, each coloring shown is `balanced':
nodes of the same color have inputs that are color-isomorphic
\cite[Section 10.3]{GS23}. The top figure corresponds to
complete synchrony, with all nodes having identical waveforms.
The bottom figure corresponds to cluster synchrony with three
clusters.

Moreover, the CPG $\{1,2,3\}$ has cyclic group symmetry $\Z_3$.
This implies that a synchrony-breaking Hopf bifurcation can create
a phase pattern in which successive nodes are phase-shifted by 
$\sot$ of the period \cite{GS02,GS23}. This synchrony
pattern corresponds to the 3-coloring in the bottom figure.
More precisely, the phase shift in a fixed 
time direction is $\pm\sot$ of the period, depending on parameters:
for simplicity we ignore the $\pm$ sign throughout.

\subsection{Motivation for Feedforward Lifts}

The structure of a feedforward lift is motivated by common
aspects of animal physiology, in which oscillations propagate along
a linear chain of neurons (or populations of neurons)
in synchrony with each other, or with specific phase relations.
This results in the longitudinal propagation of rhythmic time-periodic patterns.
Complete synchrony gives the appearance of a standing wave, and phase synchrony 
gives the appearance of a traveling wave.
Some biological examples are discussed in Section \ref{S:BM}.
Similar waves of motion are used 
to propel continuum robots for exploration \cite{MWXLW17} and medicine 
\cite{JC19,SOWRK10,ZHX20}. Synchrony patterns also occur in gene regulatory networks,
where chains of essentially identical regulatory links can act as delay lines \cite{LMRASM20}.
The states concerned are usually equilibria, but periodic oscillations can also occur \cite{PSGB10}
and phase-synchronous patterns are found, for example, in the 
repressilator, a synthetic genetic circuit that has $\Z_3$ symmetry
in an idealized model \cite{EL00}.

As in \cite{GS23,GST05,SGP03,SW23a} we adopt a strong definition of synchrony.
Two nodes are {\em synchronous} if their waveforms
(time series) are identical, and they are {\em phase-synchronous} if their waveforms
are identical except for a phase shift (time translation).
These definitions are idealizations, but they open up a
powerful mathematical approach with useful implications. Real
systems can be considered as perturbations of
idealized ones, and much of the interesting structure persists in an appropriately
approximate form; see Section \ref{S:BS}. For example,
\cite{EL00} demonstrate approximate $\sot$-period phase shifts
in a more realistic model in which the equations lack $\Z_3$ symmetry.
Many other concepts of synchrony have been studied \cite{UPLMNNS09}.

Figures \ref{F:7chainFHNsync} and \ref{F:7chainFHNZ3} of Section~\ref{S:FHNsims}
are examples of such 
signals for the network of Figure \ref{F:7nodeFFZ3}, and respectively show 
complete synchrony and a $\sot$-period phase pattern. The 
corresponding balanced colorings are those in Figure \ref{F:7nodeFFZ3}.
More generally, if the symmetry group of the CPG is a cyclic group $\Z_k$ 
of order $k$ then there can exist phase patterns with
phase shifts that are integer multiples of $T/k$, where $T$ is the period \cite{GS02,GSS88}.
These patterns are {\em rigid}, which is a form of structural stability. Specifically,
rigidity means that the phase shifts 
are preserved, as fractions of the period, after any sufficiently
small admissible perturbation of the ODE. 
Stability of these synchronous and phase-synchronous states 
depends on nonlinear terms on the ODE, usually in a complicated manner.
It also depends on the notion of stability concerned.

As remarked in \cite{SW23a}, the use of chains to propagate signals is
scarcely novel; it is simple, natural, and obvious.
However, the formal setting for network dynamics leads to rigorous proofs of some 
general model-independent results. Paramount among them are conditions for the
propagating signals to be stable. For feedforward lifts, stability decomposes
into stability against synchrony-preserving perturbations
and stability against synchrony-breaking perturbations. The first requires the
orbit of the CPG to be stable in the state space for the CPG nodes; the
second requires stability transverse to the synchrony subspace concerned.

\subsection{Stability}
There are several distinct notions of stability, even for equilibria, and
further distinctions arise for periodic orbits because of their
 global nature. The most important stability concept for a periodic orbit
comes from Floquet theory \cite{F83,HKW81}, and we use the term
`Floquet stable' for clarity. It is the stability notion adopted in this paper.

More generally, four notions of transverse stability are considered in \cite{SW23a}:
transverse Liapunov, asymptotic, and Floquet stability of the periodic orbit, and
transverse stability of the synchrony subspace. 
 Floquet stability of signals propagating along
arbitrarily long chains reduces to certain stability conditions related to the CPG
alone. Thus, if the propagating signal is stable one step along the chain, 
it remains stable for a chain of any length. A similar property holds 
for Liapunov stability if this is established using a Liapunov function,
subject to reasonable technical conditions; see \cite[Section 5.7]{SW23a}.

Transverse Floquet stability and transverse stability of the synchrony subspace are
properties of the periodic orbit on the CPG.
Transverse Liapunov and asymptotic stability are properties of 
orbits near the periodic orbit on the CPG. These concepts characterize
stability (in the corresponding sense) of the periodic orbit to synchrony-breaking
perturbations. For example, the lifted periodic orbit is Floquet stable if and 
only if the CPG orbit is Floquet stable on the CPG state space and
transversely Floquet stable for each node of the chain \cite{SW23a}.
In general, transverse Floquet multipliers must be computed numerically. 
Simulations, described below, show
 that the transverse stability condition is often satisfied in
standard models of neuron dynamics, which also tend to be 
Floquet stable. A weaker condition,
{\em transverse stability on average}, is more common, and is also
associated with Floquet stability.

A different stability measure, 
transverse stability of the synchrony subspace---either globally or on
(hence also near) the periodic orbit---occurs when the 
transverse eigenvalues of the Jacobian have negative real parts
at every point of the synchrony subspace (or just of the periodic orbit). 
This is a tempting stability criterion, because
the transverse eigenvalues of the Jacobian,
or the signs of their real parts, can often be computed analytically,
in contrast to Floquet multipliers. However, it must be treated
with care. Transverse stability in this sense implies transverse Floquet stability
for equilibria with node spaces of any dimension, 
and for periodic orbits when node spaces are $1$-dimensional. However,
the Markus-Yamabe example \cite{MY60}
shows that for node spaces of higher dimension,
this condition need not imply Floquet stability of a periodic orbit:
in this example, a nearby trajectory diverges to infinity. This
counterintuitive result disproves a conjecture that was 
widely believed at the time. 

\subsection{Biological Motivation}
\label{S:BM}

To set the scene, we begin with some typical
 examples of propagating phase-synchronous signals in biological systems. 
We discussed two such examples in \cite{SW23a} : 
peristalsis in the gut and peristalsis in the locomotion
of {\em Drosophila} larvae. Here we add three more:
the heartbeat of the medicinal leech, legged locomotion,
and locomotion of the nematode {\em Caenorhabditis elegans}.

\begin{example}\em
\label{ex:leech}
\begin{figure}[h!]
\centerline{%
\includegraphics[width=0.09\textwidth]{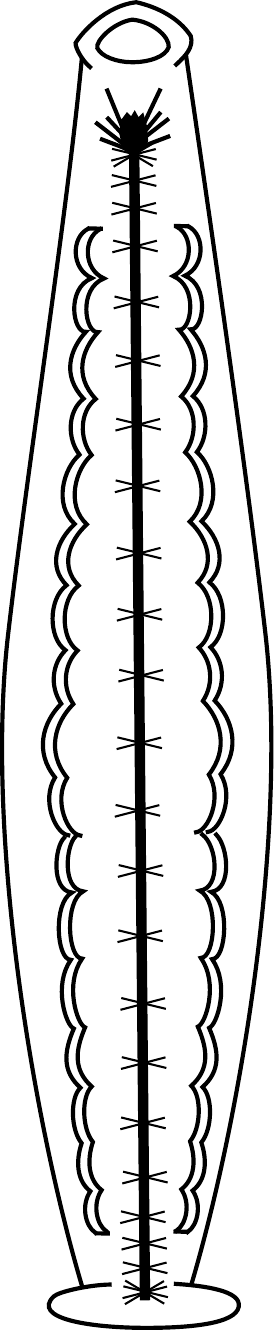}
\qquad\qquad
\includegraphics[width=0.3\textwidth]{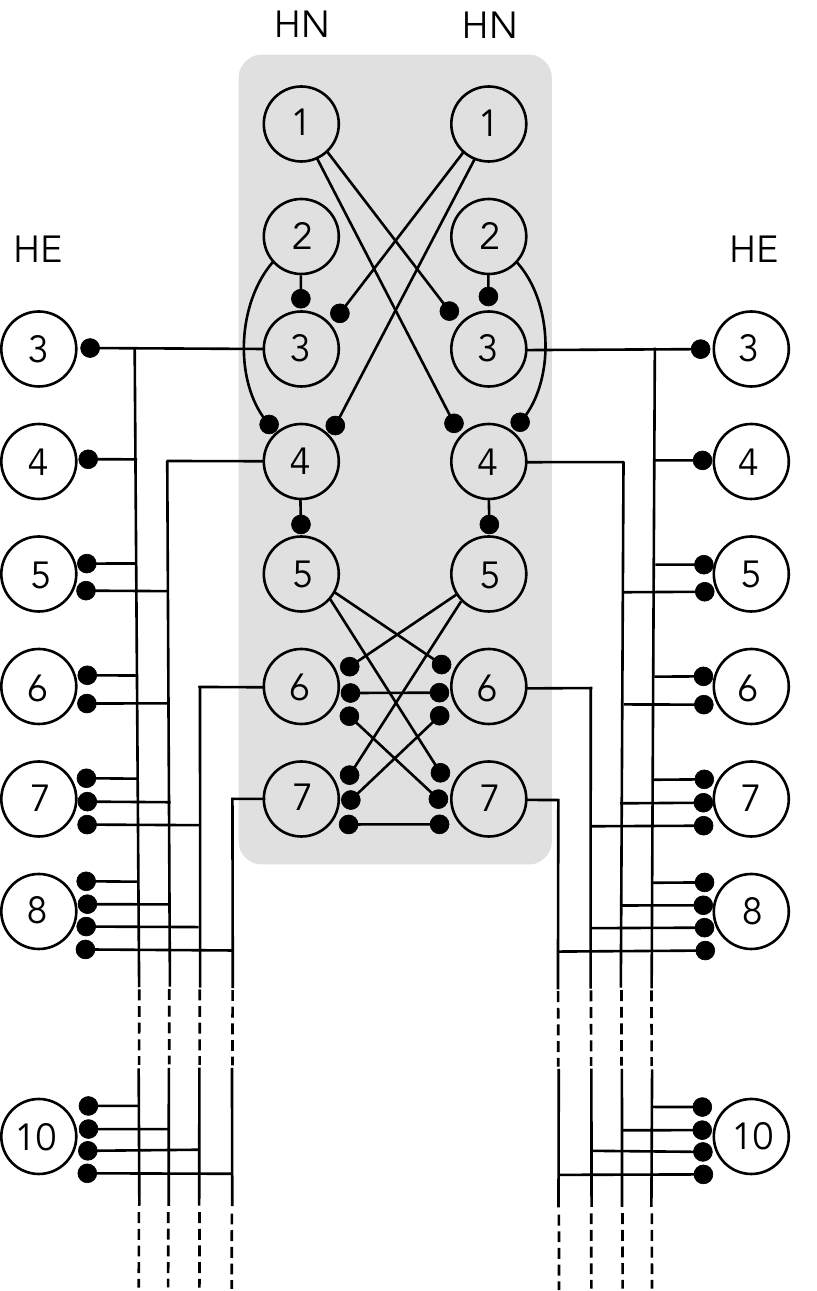}
}
\caption{{\em Left}: Medicinal leech showing two linear series of heart
chambers, one on each side, \protect redrawn from \cite{CP83}. 
{\em Right}: timing signal CPG for the leech heartbeat (shaded box)
 and propagating chains of neurons, redrawn from \protect
\cite[Figure 4A,B]{KCF05}, HE = heart excitatory motor neuron. HN = heart interneuron.
All connections shown are inhibitory.}
\label{F:timingCPG}
\end{figure}

The leech {\em Hirudo medicinalis} has two series of heart chambers,
one on each side, Figure~\ref{F:timingCPG} (left).
The chambers on one side of the animal contract in synchrony,
while those on the other side contract in a traveling wave. After
30--40 heartbeat cycles, the two sides swap patterns \cite{CP83,TS76}.
Figure~\ref{F:timingCPG} (right) shows 
an experimentally observed CPG \cite{KCF05} and a chain
(nodes 8 onwards)
along which the timing signals propagate. This chain forms
a feedforward lift of the CPG together with nodes HE 8 (left and right),
since the inputs to nodes 9 onwards are copied from node 8.

The CPG shown is a refinement of that described in \cite{CP83,CNO95}.
The architecture is essentially identical in different animals, 
except for connection strengths \cite{RNC12}.
A mathematical model that treats
the dynamics as a mode interaction is analysed in \cite{BP04}, and
an alternative model for such behaviour is proposed in \cite{PS08b}.
\end{example}

\begin{example}\em
\label{ex:locomotion}

A second class of examples is animal locomotion, especially in 
linear creatures such as arthropods, snakes, worms,
eels, and lampreys. Generally speaking, forward locomotion arises
from a  wave of muscle (hence limb, where appropriate) movements
traveling from the rear of the animal to the front. This
compresses the animal near the rear, and the compression zone moves forwards,
propelling successive segments of the body forwards.
Backward locomotion
involves a similar wave traveling in the opposite direction.

Even in animals with few limbs, the muscle and limb movements can
be modelled as traveling waves along short chains, closed up into cycles. 
These networks often exhibit synchrony or phase-synchrony,
in characteristic patterns known as {\em gaits} \cite{G74, M99}. It is widely believed that
these patterns are created by a CPG, and that in vertebrates this is
located in the spinal column \cite{G03,GE20,GJ09}. For some
organisms, such as the lamprey \cite{G03}, this has been verified directly;
see also \cite{G09,K06}. A recent survey is \cite{B19}, and the prevailing paradigm
of alternating activity in excitatory and inhibitory neurons
is laid out in \cite{AC18,MR08}. 
New observations 
challenging this paradigm are described in \cite{LPVB22}.

Gaits in bipeds, quadrupeds, and hexapods have characteristic phase patterns, 
many of which are typical in certain symmetric networks \cite{B01,CS93a,CS93b,GSBC98,GSCB99,PG06,S14}. Experiments verify that
circuits of this kind can generate basic gait rhythms \cite{CR94}, with potential
applications to legged robots \cite{IKLNPB21}. Analogous networks generate
gaits typical of myriapods such as centipedes and millipedes. Laterally related 
limbs move half a gait period out of phase (antisynchrony) in centipedes, 
and in phase in millipedes \cite[Chapter 11]{G68}. The CPG architecture suggested in \cite{GSBC98,GSCB99} predicts that
centipede locomotion should occur with wavelengths of half an odd integer
relative to the length of the organism, consistent with observations.
These phase relations between segments
are distinct from the phase relations between
extensor and flexor motions of a single limb. 
The early `half-center' architecture assumed strict alternation of extensor and flexor 
actions, but they can also partially overlap or synchronize \cite[Section 1]{MR08}.
\end{example}

The  symmetric
CPG architectures proposed in the above papers
need not be realized {\em directly} as neural circuits. Instead, a distributed network
can have a synchrony pattern such that the corresponding `quotient network' of synchronized clusters has symmetry, even though the original network has no symmetry.
A similar point is made in \cite{NVCDL13}. Indeed, general theorems in
network dynamics imply that (subject to certain technical conditions)
any structurally stable (or `rigid') phase pattern---one that 
persists after small perturbations to the model equations---should be generated
by a network that has a quotient with cyclic group symmetry; see Section \ref{S:PS}. The
`feedforward lift' construction of \cite{SW23a} at the core of this paper leads to networks of this kind,
composed of feedforward chains of simpler modules. This suggests an alternative
to the small CPG networks in the above papers, more suited to realization within the
spinal column \cite{SW23c}. 

\begin{example}\em
\label{ex:Celegans}
A widely studied example is the nematode worm
{\em Caenorhabditis elegans}, whose cellular structure is known in great detail
(see the WormAtlas~\cite{WA}). In particular, the connectome (neural wiring
diagram) is known. It includes two distinct neuronal
circuits for forward and backward movement \cite[Figures 2, 3]{MM19}, which drive muscle groups
distributed along the body of the nematode. Its motion has been modelled
using linear chains of repeating modules \cite{BBC12,IB18,OIB21,SSSIT21}.
\end{example}

Many networks in the literature have a similar repetitive feedforward structure; 
see for example \cite{MHKDA03,SMG18}.
Similar but less constrained network topology occurs in synfire chains
\cite{AHL04,ADG96,HHMD11,ZT14}. 

\begin{remark}\em
\label{R:evol}
The ability of feedforward lifts to propagate synchronous or phase-synch\-ron\-ous
signals in a natural, often stable, manner suggests a plausible, though speculative, evolutionary route
from small CPG networks to networks with long chains that automatically propagate CPG signals
stably. Namely: begin by evolving the CPG network and add a short chain. If the
signal is stable, copy the modules repeatedly to form a longer chain.
Since stability is automatically preserved, this process is guaranteed to
propagate the signals stably, low-hanging fruit for evolution, which
can readily create repetitive structures once the basic module has evolved,
as found, for example, in arthropods. 
\end{remark}

\subsection{Summary of Paper}

Section \ref{S:SNDF} is included for convenience. 
It begins with a short summary of the network dynamics formalism
of \cite{SGP03,GST05,GS23}, condensed from \cite[Section 2]{SW23a}.
It then considers feedforward lifts, the basic construction of that paper and this one,
and transverse Floquet multipliers, which determine the stability of the lifted periodic orbit
to perturbations that break synchrony or phase-synchrony.
Section \ref{S:CFM} discusses a method for
computing Floquet multipliers, which we use in subsequent numerical calculations.

Sections~\ref{S:FHNM}--\ref{S:HRM} analyse two transverse stability conditions
(transverse Floquet stability and transverse stability of the synchrony subspace)
for four common model neurons: 
FitzHugh--Nagumo, Morris--Lecar, Hodgkin--Huxley, and Hindmarsh--Rose.
For each type of model neuron we consider stability in the Floquet sense and
transverse stability of the synchrony subspace, either globally or on the periodic orbit. 
In some cases the periodic state is Floquet stable even when 
the synchrony subspace is not transversely stable, an interesting
counterpoint to the Markus--Yamabe example \cite{MY60}, which shows that
transverse stability need not imply Floquet stability.
For all of these models except Hodgkin--Huxley we briefly mention
a form of `transverse stability on average' \cite[Section 5.2]{SW23a}. 
This concept has not been made rigorous.
We analyse transverse stability analytically,
when possible, and otherwise use numerics. For simplicity
we study the `toy' network of Figure \ref{F:7nodeFFZ3}, which has
a very simple GPG that robustly generates oscillations with a regular
phase pattern, and a long enough chain to illustrate the theory.

Section \ref{S:BS} addresses the issue whether the propagating
signals persist after synchrony-breaking modifications
or perturbations of the model. This is important for applications,
since the conditions assumed in an idealized feedforward chain,
and the strict definition of synchrony, apply only approximately 
in real systems. We give qualitative theoretical reasons to suppose that
the results are reasonably robust, and support them by quantitative
numerical simulations.

Finally, we summarize the main conclusions in Section~\ref{S:C}.

The numerical calculations in Sections \ref{S:FHNM}--\ref{S:HRM} 
were carried out using Mathematica. Those in Section \ref{S:BS}
were performed in MatLab.

\section{Summary of Network Dynamics Formalism}
\label{S:SNDF}

We work in a formal framework for analysing networks of coupled
dynamical systems (ODEs) \cite{GS23,GST05, SGP03}. 
Such systems are represented by directed graphs
in which nodes and directed edges (`arrows') are labelled
with `types'. (In graph theory these are called {\em colored graphs} \cite{W61} or {\em multilayered graphs} \cite{HK20}, but
we reserve the term `color' to define sets of synchronous nodes.) 
Nodes of the same type have the same state space,
and arrows of the same type represent identical types of coupling.
Nodes with isomorphic sets of input arrows obey identical ODEs
when corresponding couplings are identified. We outlined this formalism
in \cite[Section 2]{SW23a}; it is a natural generalization of
the formalism for equivariant dynamical systems \cite{GSS88}
when applied to symmetric networks.
We restate a few basic concepts here for convenience.

\subsubsection{Admissible ODEs}
Each network diagram is associated with a class of {\em admissible ODEs}, which
are the most general ODEs that respect the network structure.
Number the nodes from $1$ to $n$ and let $x_c$ be the state variable
for node $c$. We assume that $x_c \in \R^{n_c}$, a finite-dimensional real vector space.
(Other choices are possible; in particular the circle $\Sone$ applies
to phase oscillator networks \cite{K84}.) The overall state of the system is then
$x = (x_1,\ldots, x_n)$. The dynamics takes the form
\begin{equation}
\label{E:gen_admiss_ODE}
\dot x_c = f_c(x_c, x_{i_1}, \ldots, x_{i_k})	\qquad 1 \leq c \leq n
\end{equation}
where  $i_1, \ldots, i_k$ are the tail nodes of arrows that input to node $c$.

\begin{remark}\em
In this approach to network dynamics, an output arrow comes
into play only as an input to its head node, a convention sometimes referred to as
the {\em influence network} of the model. In some areas of application,
such as biochemical networks and epidemiology, different conventions
are standard \cite[Examples 2.2--2.4]{GS23}. ODEs constructed
using those conventions remain admissible for the influence network
and therefore fall into the scope of that formalism. However,
unlike many of these conventions,
arrows in the network diagram do not correspond to specific {\em terms}
in the associated model ODEs, and the formalism does not (as sometimes
assumed) include only two-body interactions, as equation \eqref{E:gen_admiss_ODE}
clearly demonstrates. In fact, the network diagram in this formalism can best be
viewed as a very general type of hypergraph \cite{B73}, in which bunches of
input arrows of the same type correspond to hyperarcs satisfying certain
symmetry constraints. These hyperarcs determine the functional form of the ODE,
not specific terms in its component functions \cite[Section 9.3.3]{MBSS25}.
\end{remark}

For convenience we assume that $f_c$ is of class $C^\infty$ (infinitely differentiable),
but less stringent smoothness conditions are possible with no change to proofs.
Nodes of the same `input-type' have the same function $f_c$,
applied to their respective sets of inputs. (The same node may 
send multiple inputs, so these concepts are formalized in terms of the input arrows.)
For example, the admissible ODEs for Figure~\ref{F:7nodeFFZ3} have the following form:
\begin{equation}
\label{E:7nodeFFZ3}
\begin{array}{rcl}
\dot{x}_1 &=& f(x_1,x_3) \\
\dot{x}_2 &=& f(x_2,x_1) \\
\dot{x}_3 &=& f(x_3,x_2) \\
\dot{x}_4 &=& f(x_4,x_3) \\
\dot{x}_5 &=& f(x_5,x_4) \\
\dot{x}_6 &=& f(x_6,x_5) \\
\dot{x}_7 &=& f(x_7,x_6) 
\end{array}
\end{equation}
So, for example, $f_1(x_1,x_3)=f(x_1,x_3)$, and so on. These variables
need not be, and often are not, $1$-dimensional.
The same function $f$ is used to specify all components, because all 
nodes have the same node-type and all arrows
have the same arrow-type. The domains of these copies $f_i$ of $f$ are
technically different, being Cartesian products of state spaces of the respective nodes,
but they are canonically identified with the domain of $f$.
By convention, the first variable in $f$ is the node variable;
the other variables correspond to
the input nodes. In this case, the input node is unique so there is only one other variable. 
See \cite[Example 2.2]{SW23a} for full details and caveats.

\subsubsection{Colorings, Synchrony, and Quotient Networks}

Synchrony relations are determined by a {\em coloring} $\kappa$
of the nodes, associating to each node $c$ a {\em color},
which is an element $\kappa(c)$ of some specified
set $\KK$. Synchronous nodes are assigned the same color, so
$c, d$ being synchronous means that $\kappa(c) = \kappa (d)$. The key concept
here is that of a {\em balanced} coloring. Balance requires nodes of the same color
to have input sets with matching colors, in a manner that preserves
node- and arrow-types. The coloring in Figure~\ref{F:7nodeFFZ3}
is balanced, because all nodes have the same input-type, 
every white node receives one black input, 
every grey node receives one white input, 
and every black node receives one grey input.

The corresponding {\em synchrony subspace} 
\[
\Delta_\kappa = \{ x : \kappa(c)=\kappa(d) \implies x_c = x_d \}
\]
corresponds to states with the given synchrony pattern. Balance implies that
$\Delta_\kappa$ is invariant under any admissible ODE. 
Conversely, any subspace invariant under every admissible ODE
is equal to $\Delta_\kappa$ for some balanced coloring $\kappa$.
The restriction of the ODE to this subspace determines the dynamics of the 
synchronized {\em clusters}, where a cluster is the set of all nodes
with a given color. Identifying the nodes in each cluster,
and copying arrows according to the colors of their heads and tails,
leads to the {\em quotient network} for the coloring. For example,
the quotient for the 3-coloring in Figure \ref{F:7nodeFFZ3} (bottom)
is a 3-node directed ring. In this case it is isomorphic to the CPG $\{1,2,3\}$,
but in general a quotient network need not be a subnetwork.

Any admissible ODE
on the full network restricts to one on the quotient network; conversely
any admissible ODE
on the quotient network {\em lifts} to one on the full network.
Synchronous or phase-synchronous solutions restrict and lift in the 
same manner, preserving synchrony and phase synchrony.

\subsubsection{Feedforward Lift}

Let $\GG$ be any network. A {\em feedforward lift} of $\GG$
is a network $\widetilde{\GG}$ such that $\GG$ is both a subnetwork
and a quotient network of $\widetilde{\GG}$ by some balanced coloring, and all arrows of $\widetilde\GG$
other than those in $\GG$ are feedforward. That is, we can order the nodes
not in $\GG$ so that the tail $\tl(e)$ of every arrow $e$ is either in $\GG$, 
or is less than the head $\hd(e)$ in that ordering \cite[Theorem 4.11]{GS23}. Moreover,
every node in $\widetilde\GG$ has the same color as one in $\GG$.
Again Figure~\ref{F:7nodeFFZ3}
is an example, with a suitable ordering indicated by the numbers. Here $\GG$
is the initial CPG with nodes $\{1,2,3\}$ and the three arrows between them, 
which form a cycle.

As in \cite{SW23a}, the central topic is a periodic orbit $A=\{a(t)\}$ for 
some $\GG$-admissible ODE, 
lifted to a periodic orbit $\tilde A=\{\tilde a(t)\}$ for $\widetilde\GG$. 
Any synchrony or phase pattern on $A$ lifts to one on $\tilde A$. 

\subsubsection{Stability and Transverse Floquet Multipliers}

If $A$ is stable on the state space of the CPG nodes, then the lift
$\tilde A$ is stable to {\em synchrony-preserving} perturbations
within the synchrony subspace.
However, $\tilde A$ might be unstable
to {\em synchrony-breaking} perturbations, transverse to the synchrony 
subspace. We therefore investigate its stability, considering two main senses
of that word.
The feedforward structure
of the network implies that the Jacobian (linearization) of the ODE is block-triangular;
the first block corresponds to the CPG and subsequent blocks correspond 
to nodes running through the chain in feedforward order \cite[Corollary 3.8]{SW23a}.
Thus its eigenvalues are those of the diagonal blocks. 
For a periodic orbit, an analogous result applies to Floquet stability:
the Floquet operator
is (conjugate to) a block-triangular matrix, so the Floquet multipliers
are those of the individual blocks \cite[Section 5.1]{SW23a}. The {\em transverse Floquet multipliers}
for $A$ come from all blocks except that for the CPG. The main result
states that a necessary and sufficient condition for 
 $\tilde A$ to be Floquet stable is that $A$ is Floquet stable and all transverse
 Floquet multipliers have absolute value less than $1$  \cite[Theorem 4.4]{SW23a}.
 
 In more detail, we need:
 
 \begin{definition}\em
\label{D:TFE}
Let $f_c$ be the $c$th component of an admissible ODE,
$[c]$ a node in the CPG with the same color as node $c$,
and $\{a(t)\}$ a periodic orbit of the CPG that is lifted to
a periodic orbit $\{\tilde a(t)\}$ for the full network. Denote
the partial derivative with respect to $x_{[c]}$ by $\Dr_{[c]}$. Then:

(a)
With the above notation, the {\em transverse Floquet equation} of $A$,
for node $c \in \CC^*$ is
\begin{equation}
\label{E:TFE}
\dot y_c = \Dr_{[c]}f_{[c]}|_{a(t)} y_c
\end{equation}
Here 
Equation \eqref{D:TFE} depends only on the ODE for the CPG and the periodic orbit
$\{a(t)\}$. This is why we refer to the transverse Floquet equation of $A$,
rather than of $\tilde{A}$.

(b)
By Floquet theory \cite{HKW81}, every solution of \eqref{D:TFE} has the form
\[
y_c(t) = P_c(t)\ee^{B_ct}v
\]
for a constant vector $v$. Here $P_c(t)$ is $T$-periodic and $B_c$ is a constant matrix. 
The matrix $M_c = \ee^{B_cT}$ is the {\em transverse Floquet matrix}  for node $c$.

(c)
The matrix $B_c$ is the {\em transverse Floquet exponent matrix} for node $c$.

(d)
The periodic orbit $A=\{a(t)\}$ is {\em transversely Floquet stable} at node $c$ if all eigenvalues
of $M_c$ have absolute value $<1$.
Equivalently, all eigenvalues of $B_c$ have negative real part.

(e) The eigenvalues of $M_c$ are the {\em transverse Floquet multipliers} of $A$ for node $c$.

(f) The eigenvalues of $B_c$ are the {\em transverse Floquet exponents} of $A$ for node $c$.
\end{definition}

 One main result of \cite{SW23a} is:
 
 \begin{theorem}
\label{T:FFStab}
Let $\{\tilde a(t)\}$ be a feedforward lift of the periodic orbit $\{a(t)\}$
on $P_\CC$. Then:

{\rm (a)} The Floquet multipliers for $\{\tilde a(t)\}$ are the
Floquet multipliers for $a(t)$, together with the transverse Floquet multipliers of $a(t)$
for all $c \in \CC^*$.

{\rm (b)} The transverse Floquet multipliers for $c \in \CC^*$ are the
same as those for $[c] \in \CC$.

{\rm (c)} $\{\tilde a(t)\}$ is stable on $P$ if and only if
$\{a(t)\}$ is stable on $P_\CC$ and, for
 all nodes in $\CC$, all transverse Floquet multipliers of $a(t)$
have absolute value $<1$. 
\end{theorem}

The Floquet multipliers for $a(t)$ always include an eigenvalue $1$
corresponding to the periodic orbit \cite[Chapter 1 Section 4]{HKW81}. This is already present for the lifted orbit
$\{\tilde a(t)\}$, so the transverse Floquet multipliers generically do not
include $1$.

 \begin{remark}\em
It is well-known that symmetry can create  
multiple eigenvalues generically \cite[Chapter XII Section 2]{GSS88},
and this applies in particular to symmetric networks. In network dynamics, 
other aspects of network topology can also create multiple eigenvalues,
and feedforward lifts provide examples of this phenomenon.
Conditions (a, b) imply that long chains give rise to Floquet multipliers of high
multiplicity, even when the overall network $\widetilde \GG$ has trivial symmetry.
Analogous effects for critical eigenvalues at equilibria are well known; 
see \cite[Example 11.1]{GS06} 
and \cite[Section 21.1]{GS23}. High multiplicity of transverse critical eigenvalues
implies that synchrony-{\em breaking} bifurcations for these networks can be very complex.
\end{remark}

 \section{Phase Synchrony}
 \label{S:PS}
 If the CPG has cyclic group symmetry, suitable admissible ODEs naturally support
discrete rotating waves, with phase shifts related to a generator of the cyclic group
\cite{GS02,GSS88,SP08}.
Such phase patterns are {\em rigid}. A hyperbolic periodic orbit 
persists after any $C^1$-small perturbation of the ODE. Rigidity means that the
phase pattern also persists, with phase shifts equal to the same fractions of the period.
Conversely, subject to certain technical conditions, rigid phase patterns
conjecturally imply cyclic symmetry on the quotient network by
the synchrony coloring. This conjecture has been proved for a broad class of
networks~\cite{GRW10,GRW12} (the results are stated there for all networks, but
a technical condition was overlooked, see \cite[Section 1.4]{S22}
or \cite[Section 15.1]{GS23}). It has also been proved in full generality
under hypotheses slightly stronger than hyperbolicity \cite{S22}.

\begin{figure}[h!]
\centerline{%
\includegraphics[width=0.35\textwidth]{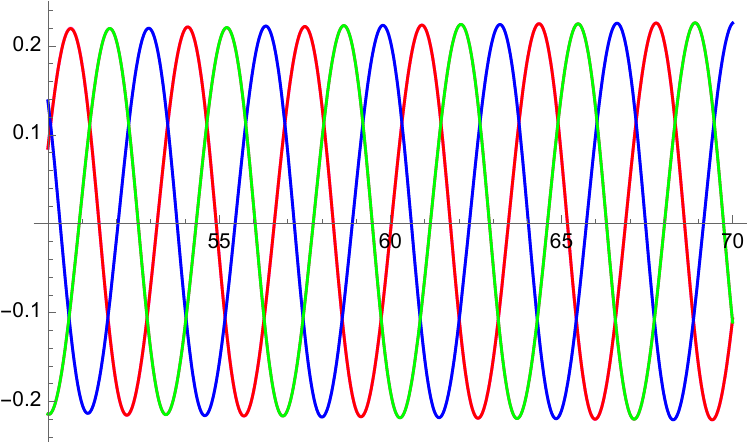} \qquad
\includegraphics[width=0.35\textwidth]{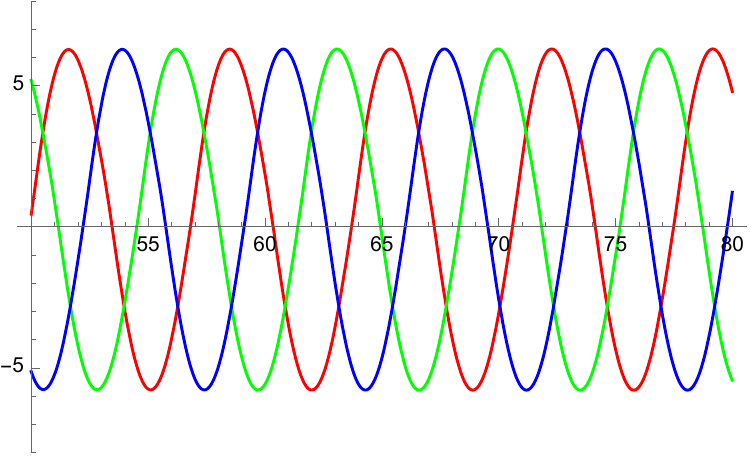} 
}
\centerline{%
\includegraphics[width=0.35\textwidth]{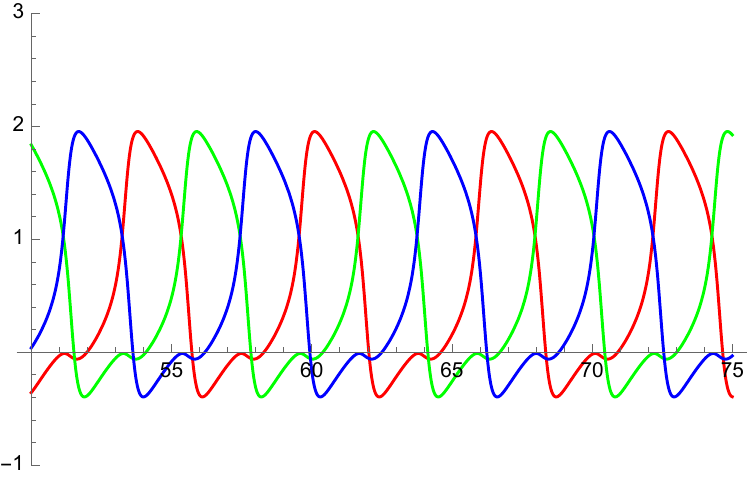} \qquad
\includegraphics[width=0.35\textwidth]{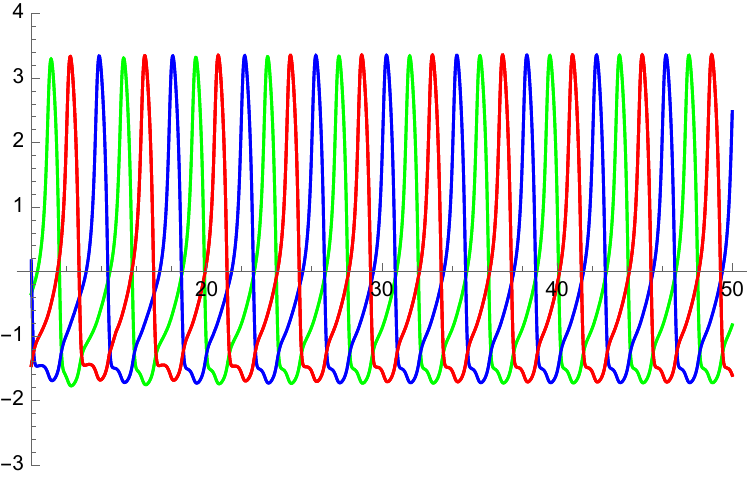}
}
\caption{$\Z_3$ traveling wave periodic states in a 3-node ring,
for four standard model neurons. Each subfigure shows three
superposed time series of corresponding node variables. {\em Top left}: FitzHugh--Nagumo.
{\em Top right}: Morris--Lecar. {\em Bottom left}: 
Hodgkin--Huxley. {\em Bottom right}: Hindmarsh--Rose.
}
\label{F:3-waves}
\end{figure}

For example, the CPG of Figure~\ref{F:7nodeFFZ3}, a cycle of length 3,
has $\Z_3$ symmetry. It supports periodic states where phases of successive nodes
are shifted by $\sot$ of the period. Figure \ref{F:3-waves} shows superposed
numerically computed examples of time series for nodes $1,2,3$, for
FitzHugh--Nagumo, Morris--Lecar, 
Hodgkin--Huxley, and Hindmarsh--Rose models. 
For details see the corresponding sections below, where the same
figures appear as part of a more general investigation.)

When lifted to $\widetilde\GG$
such orbits resemble traveling waves. The theory generalizes to
these phase patterns. 
Indeed, phase relations not caused by symmetries are also 
copied along the chain, and similar remarks apply, although we do not
pursue this possibility here.
Subject to some minor technical conditions,
the setting is as follows; see \cite[Section 6]{SW23a} for details.

Let $\GG$ be a CPG with nodes $\{1 \ldots, k\}$ and $\Z_k$ symmetry
generated by a permutation $\alpha$, which cycles the
nodes in the CPG so that $\alpha(1)= 2, \alpha(2)=3$ \ldots, $\alpha(k)=1$.
That is, $\alpha = (1\, 2\, 3\, \ldots\, k)$ in cycle notation.
Choose an admissible map $f$ so that the ODE $\dot x = f(x)$
has a discrete rotating wave state $u(t)$ satisfying the phase relation
$\alpha u(t) = u(t+rT/k)$, where $r$ is an integer and $0 \leq r \leq k-1$.
Choose a {\em module} $\MM$ whose nodes are a set of representatives
of the $\Z_k$-orbits. Assign phase $0$ to these nodes, so that the other
 $\Z_k$-orbits correspond to phase shifts $rT/k, 2rT/k, \ldots, (k-1)rT/k$.
 Copy the module (along with any arrows whose heads and tails lie in the module)
 to obtain $\MM_{k+1}, \MM_{k+2}, \ldots, \MM_l$. 
Assign phases $0, rT/k, \ldots, (l-k)rT/k$ to nodes in  $\MM_{k+1}, \MM_{k+2}, \ldots, \MM_l$.
Assign input arrows to these nodes, preserving the arrow type and the phase
relations in $\GG$. Do so in a manner that makes all new arrows feedforward.
Finally, rewire internal arrows in $\MM_{k+1}, \MM_{k+2}, \ldots, \MM_l$,
preserving the arrow-type and the phase
relations in $\GG$. Do so in a manner that makes all rewired arrows feedforward.

In this context, assuming the above notation, the following result is proved
in \cite[Theorem 6.1]{SW23a}:

\begin{theorem}
\label{T:TWtranseigen}
Assume that $\GG$ has nodes $\CC = \{1, \ldots, m\}$ 
with a cyclic automorphism group $\Z_k = \langle \alpha \rangle$,
such that $n=mk$ and $\alpha$ acts like the cycle $(1\, 2\, \ldots\, k)$
on all of its orbits on $\GG$. Let $u(t)$ be a $T$-periodic solution of an admissible ODE
with a discrete rotating wave phase pattern induced by the symmetry,
and let the module $\MM$ be a set of orbit representatives.
Let $\widetilde{\GG}$ be obtained
by lifting appropriate copies of translates of this module by $\Z_k$. Then 

{\rm (a)} The periodic state $u(t)$ on $\GG$ lifts to a $T$-periodic traveling wave
state $\tilde{u}(t)$ for $\widetilde{\GG}$ with phases corresponding to the extra copies
$\MM_{k+1}, \MM_{k+2}, \ldots, \MM_l$ of $\MM$.

{\rm(b)}
The Floquet exponents (evaluated at any point) are 
those on the module $\MM$, together with the transverse
Floquet exponents for $\MM$.

{\rm(c)} If the Floquet exponents  and
the transverse Floquet exponents on $\MM$ have negative real part,
then $\tilde{u}(t)$ is Floquet stable.
\end{theorem}

In particular, the transverse Floquet multipliers need be computed only for nodes in
the module $\MM$, however large the feedforward lift $\widetilde{\GG}$ may be.
This remark applies to any periodic state of the CPG, and does not require
phase relations to be caused by symmetry.

\subsubsection{Comparison of Stability Concepts}
 
 Floquet multipliers generally cannot be computed analytically, except when
 node spaces are 1-dimensional. A more tractable, but less rigorous, stability
 condition is {\em transverse stability} of $\tilde A$ relative to 
 the synchrony subspace $\Delta_\kappa$. This requires all transverse
 eigenvalues of the Jacobian---those not coming from the CPG---to have negative
 real part when evaluated at any point of $A$. When node spaces are 1-dimensional
 this is equivalent to the transverse
 Floquet multipliers having absolute value less than $1$, but this
 result can be false for higher-dimensional node spaces \cite[Examples 5.3, 5.4]{SW23a}.
 
 The main focus of this paper is to compare the behaviour of
transverse Floquet multipliers and transverse eigenvalues of 
 the Jacobian, for a variety of standard neuron equations---namely, Fitzhugh--Nagumo,
Morris--Lecar, Hodgkin--Huxley, and Hindmarsh--Rose neurons.
 To standardize the method we mainly do this for the network of Figure~\ref{F:7nodeFFZ3}. 
 We also discuss Floquet stability of the CPG orbit $A$ for Fitzhugh--Nagumo neurons.
Pragmatically, if numerical simulations produce traveling waves
with the $\sot$-period phase pattern for randomly chosen initial conditions,
the periodic orbit concerned should be stable. Otherwise this solution would
not be observed. Thus the Floquet stability results below should not
come as a surprise. 

\subsection{Computation of Floquet Multipliers}
\label{S:CFM}

We recall the 
classical stability criterion of Floquet theory \cite[Chapter 1 Section 4]{HKW81}
and describe how we compute Floquet multipliers.
Linearizing the ODE about the periodic orbit $A$ gives a non-autonomous ODE
\begin{equation}
\label{E:Lin}
\dot y = M(t)y
\end{equation}
 where $M(t) = \Dr_yf |_{a(t)}$
is $T$-periodic. A {\em fundamental matrix} for \eqref{E:Lin} is a time-dependent
 matrix $Y(t)$ such that any solution $y(t)$ has the form
 $y(t) = Y(t)v$ 
 for some constant vector $v$.
Floquet's Theorem \cite{F83} implies the
existence of a fundamental matrix. Moreover,
there is a $T$-periodic matrix function $P(t)$ and a matrix $B$
 such that every fundamental matrix can be written as
\begin{equation}
\label{E:compFloq}
Y(t) = P(t)\ee^{Bt}
\end{equation}
The eigenvalues $\beta_i$ of $B$ are {\em Floquet exponents}
and the eigenvalues $\rho_i$ of $\ee^{BT}$ are {\em Floquet multipliers}.
The condition for Floquet stability is that all Floquet exponents $\beta_i$ have negative real part,
except for a simple eigenvalue $0$ corresponding to the orbit.
Equivalently, all $\rho_i$ are in the interior of the unit circle, except for
a single eigenvalue $1$. In modern dynamical systems theory the
matrix $\ee^{BT}$ is replaced by the Jacobian of a Poincar\'e map
at the fixed point determined by the periodic orbit \cite[Section 1.5]{GH83}. This reduces the
dimension by one, and
excludes the eigenvalue $1$ whose eigenvector is tangent to the periodic orbit.

It is well known that Floquet exponents/multipliers can seldom be
found analytically. However, numerical calculations are straightforward
for specific examples. In Sections \ref{S:FHNM}--\ref{S:HRM} we present numerical results
on some Floquet multipliers. The corresponding Floquet exponents
can be found by taking logarithms. 

Our method is based on \eqref{E:compFloq}. This implies that,
for any solution $Y(t)$, 
\beqn
Y(t_0) &=& P(t_0)\ee^{Bt_0}\\
Y(t_0+T) &=& P(t_0+T)\ee^{B(t_0+T)} = P(t_0)\ee^{BT}\ee^{Bt_0}
\eeqn
since $Bt_0$ and $BT$ commute. Therefore
\beqn
Y(t_0+T) &=& P(t_0)\ee^{BT}\ee^{Bt_0} \\
	&=& (P(t_0)\ee^{BT})((t_0)^{-1} Y(t_0)) \\
	&=& [P(t_0)\ee^{BT}P(t_0)^{-1}]Y(t_0)
\eeqn
The matrix $E=P(t_0)\ee^{BT}P(t_0)^{-1}$ is conjugate to $\ee^{BT}$,
so it has the same eigenvalues. These are the required Floquet multipliers.

The period $T$ can be computed numerically. Having done so,
one way to compute $E$ is to take $n$ linearly independent
initial conditions $Y_i(0)$, such as the column vectors $\mathbf{e}_i$
with $1$ in the $i$th place and $0$ everywhere else. Compute
$Y(t_0)$ and $Y(t_0+T)$ numerically. In each case these
sets of vectors are also linearly independent, so we can solve
the system
\[
E Y_i(t_0) = Y_i(t_0+T)\quad 1 \leq i \leq n
\]
for $E$, whose eigenvalues are the Floquet multipliers.
Explicitly, consider the $Y_i$ as column vectors and let
\[
U = [Y_1(t_0)| \cdots |Y_n(t_0)]\qquad V = [Y_1(t_0+T)| \cdots |Y_n(t_0+T)]
\]
be the matrices obtained by concatenating the indicated columns.
Then $EU=V$, so $E = VU^{-1}$.

This approach lets us start the integration from time $0$ and then
choose $t_0$ large enough for the initial state to have settled down
close to the periodic cycle, so the computed $a(t)$ effectively lie on the cycle.
These values can then be fed into \eqref{E:Lin}. This method yields both 
the Floquet multipliers for the periodic orbit $A$ and the transverse Floquet multipliers
for $\widetilde{A}$. 

Some care must be taken to avoid numerical instabilities, and to ensure that
the computed oscillation has settled down sufficiently close to the actual periodic orbit.

\begin{remark}\em
For a specific small network such as the 7-node chain, 
both the Floquet multipliers for the cycle $A$ and the transverse
Floquet multipliers can be found directly,
by computing the the Floquet multipliers for the  lifted periodic orbit. 
Theorems~\ref{T:FFStab} and \ref{T:TWtranseigen} imply that
the same computation gives the stability of 
$\tilde a(t)$ for {\em any} feedforward lift $\widetilde\GG$, however large it may be.
\end{remark}

\section{FitzHugh--Nagumo Model}
\label{S:FHNM}

We now examine two main notions of transverse stability for
feedforward lifts whose nodes are standard model neurons,
beginning with FitzHugh--Nagumo neurons. These stability notions are
transverse Floquet stability and transverse stability of the synchrony subspace.
We also consider Floquet stability of the CPG periodic orbit $A$ for 
FitzHugh--Nagumo neurons. All time series and Floquet multipliers 
in Sections \ref{S:FHNM} -- \ref{S:HRM} are computed using Mathematica.

For all of the standard models, we use the same running example as \cite{SW23a},
namely Figure~\ref{F:7nodeFFZ3}. This 7-node network is
a feedforward chain with a single feedback connection from
node 3 to node 1. Nodes $\{1,2,3\}$ and connecting
arrows are interpreted as a CPG with $\Z_3$ symmetry, 
and the rest of the network is the chain in a feedforward lift. The `colors' of the nodes 
(white, grey, black) indicate a synchrony pattern that is preserved by the lift. They
also correspond to synchronous nodes in a phase pattern with $\sot$-period 
relative phase shifts.

The FitzHugh--Nagumo model of a single neuron~\cite{F61, NAY62} takes the form
\begin{eqnarray}
\label{E:FHN1}
\dot{V} &=& V(a-V)(V-1)-W+I \\
\label{E:FHN2}
\dot{W} &=& bV-\gamma W 
\end{eqnarray}
where:

\quad $V$ is the membrane potential

\quad $W$ is a recovery variable

\quad $I$ is the magnitude of stimulus current

\quad $0<a<1$ is a parameter

\quad $b, \gamma >0$ are parameters.  

\noindent
In principle, the voltage $V$ and input $I$ 
can have either sign.

The transverse Floquet equations for FitzHugh--Nagumo neurons
are derived explicitly in \cite[Example 5.3]{SW23a}. We discuss the results
for the special case of Figure~\ref{F:7nodeFFZ3}, but since the transverse Floquet equations
depend only on the CPG and its periodic orbit $\{a(t)\}$, the same 
results apply to any feedforward lift,
such as a linear chain of $n$ FitzHugh--Nagumo neurons for any $n$, 
or a branching tree of such chains.

\subsection{Simulations}
\label{S:FHNsims}

We model Figure~\ref{F:7nodeFFZ3} when
the nodes are FitzHugh--Nagumo neurons with voltage coupling.
By \cite[Theorem \ref{T:TWtranseigen}]{SW23a} 
it suffices to calculate the transverse
eigenvalues and transverse Floquet multipliers 
at a single node, which plays the role of the module $\MM$.

Let
\beqn
G(V,W,I) &=& V (a - V) (V - 1) - W + I \\
H(V,W) &=& bV - \gamma W
\eeqn
Model equations are:
\begin{equation}
\label{E:7nodeFHN}
\begin{array}{rcl}
\dot{V}_1 &=& G(V_1,W_1,I)+cV_3 \qquad \dot{W}_1 = H(V_1,W_1) \\
\dot{V}_2 &=& G(V_2,W_2,I)+cV_1 \qquad \dot{W}_2 = H(V_2,W_2)\\
\dot{V}_3 &=& G(V_3,W_3,I)+cV_2 \qquad \dot{W}_3 = H(V_3,W_3)\\
\dot{V}_4 &=& G(V_4,W_4,I)+cV_3 \qquad \dot{W}_4 = H(V_4,W_4)\\
\dot{V}_5 &=& G(V_5,W_5,I)+cV_4 \qquad \dot{W}_5 = H(V_5,W_5)\\
\dot{V}_6 &=& G(V_6,W_6,I)+cV_5 \qquad \dot{W}_6 = H(V_6,W_6)\\
\dot{V}_7 &=& G(V_7,W_7,I)+cV_6 \qquad \dot{W}_7 = H(V_7,W_7)
\end{array}
\end{equation}
Here $c$ is the strength of the voltage coupling,  $V_c$ is the voltage at node $c$, 
and $W_c$ is the recovery variable at node $c$.

\begin{figure}[h!]
\centerline{%
\includegraphics[width=0.2\textwidth]{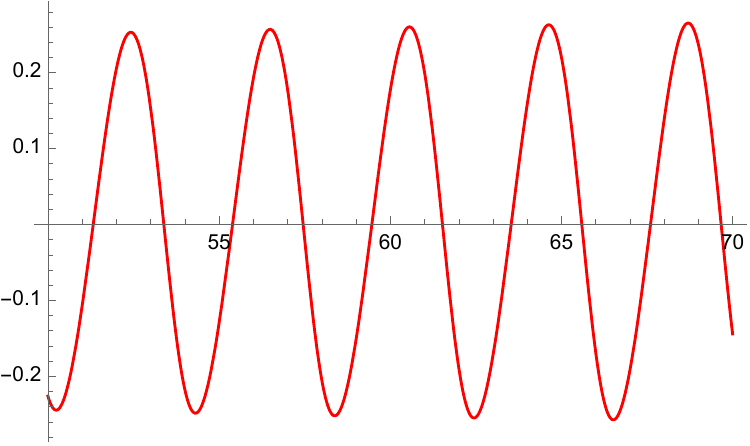} \quad
\includegraphics[width=0.2\textwidth]{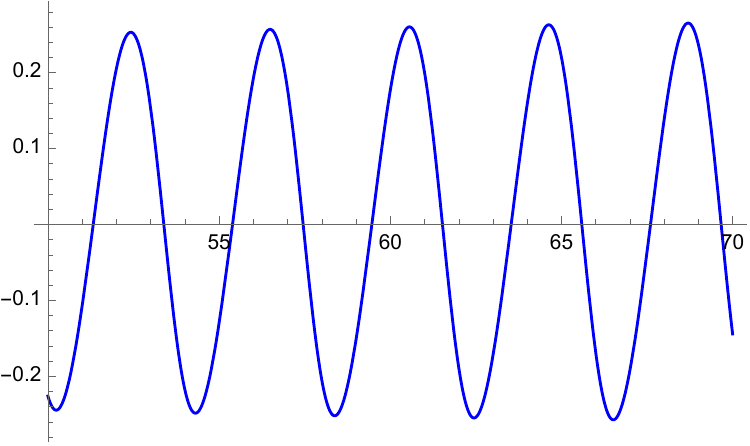} \quad
\includegraphics[width=0.2\textwidth]{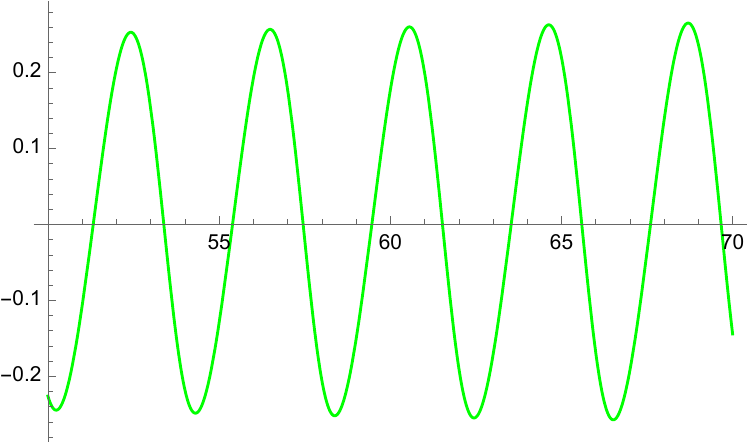} \quad
\includegraphics[width=0.2\textwidth]{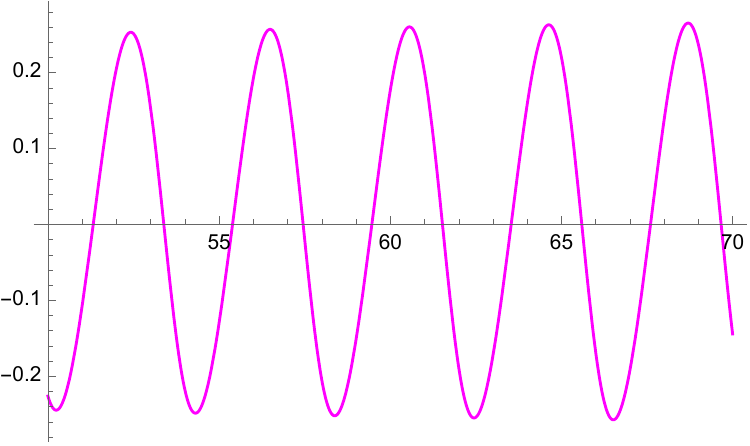} \quad
}
\vspace{.1in}
\centerline{%
\includegraphics[width=0.2\textwidth]{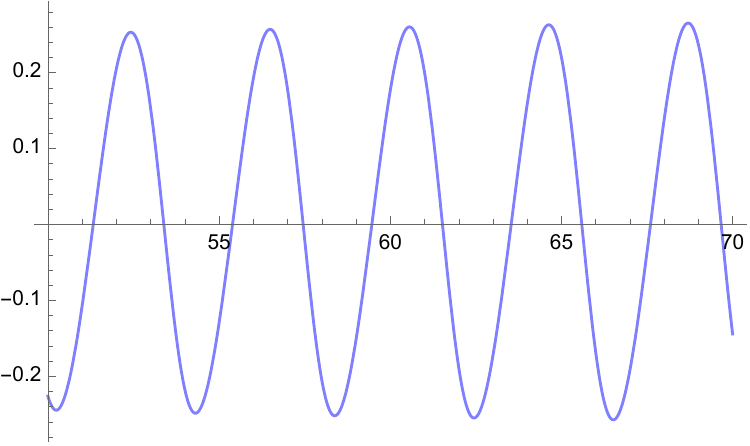} \quad
\includegraphics[width=0.2\textwidth]{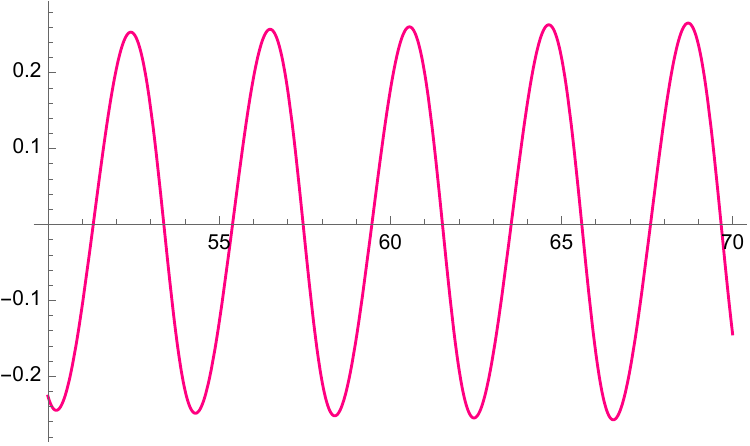} \quad
\includegraphics[width=0.2\textwidth]{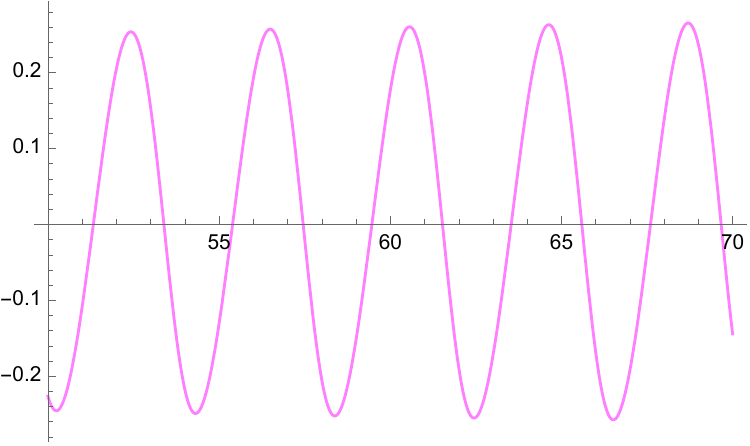} \quad
\includegraphics[width=0.2\textwidth]{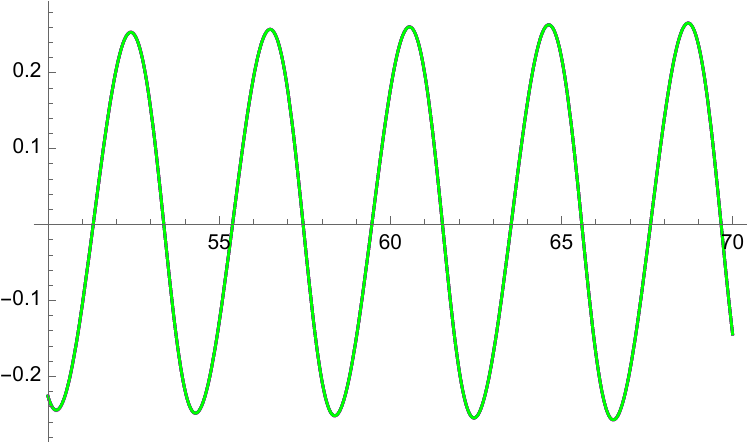} \quad
}
\caption{Synchronous periodic state. Time range = $[50,70]$ to eliminate transients.
{\em Top left to bottom right}: Nodes $1$--$7$; all nodes superposed.}
\label{F:7chainFHNsync}
\end{figure}

Figure~\ref{F:7chainFHNsync} shows a synchronous periodic state.
Parameters are:
\begin{equation}
\label{E:FHNparams1}
I=0 \quad a =0.05 \quad b = 2.5 \quad \gamma=0.3 \quad c = 0.4
\end{equation}

\begin{figure}[h!]
\centerline{%
\includegraphics[width=0.2\textwidth]{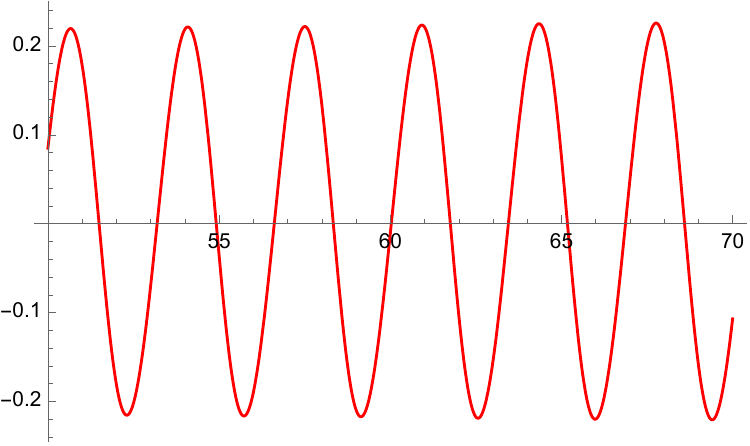} \quad
\includegraphics[width=0.2\textwidth]{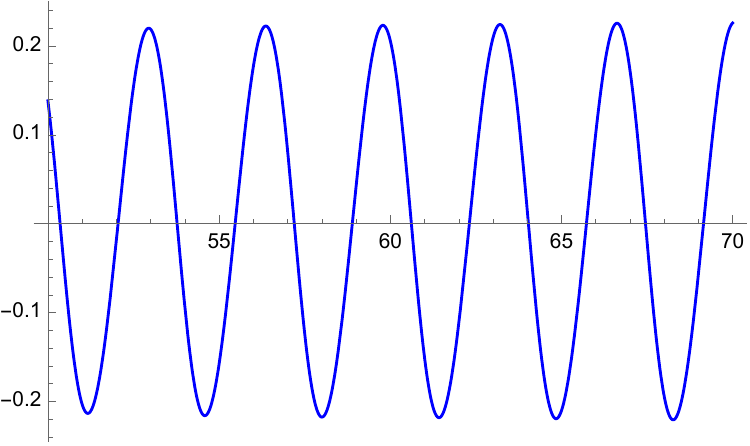} \quad
\includegraphics[width=0.2\textwidth]{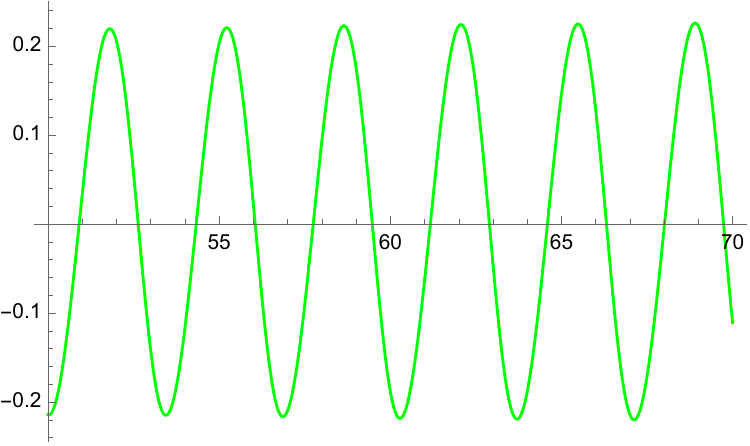} \quad
\includegraphics[width=0.2\textwidth]{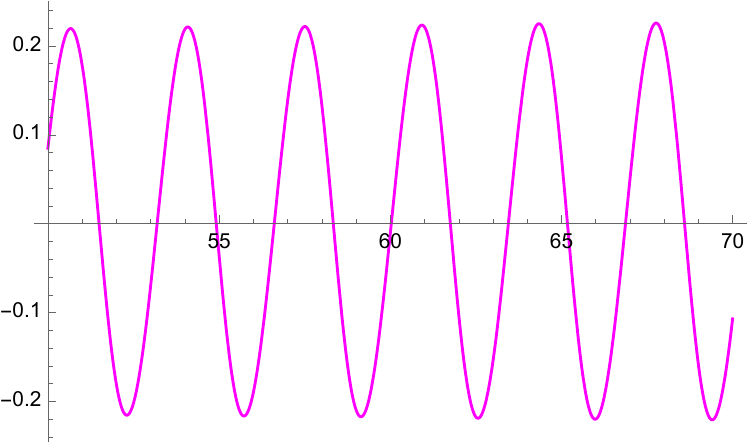} \quad
}
\vspace{.1in}
\centerline{%
\includegraphics[width=0.2\textwidth]{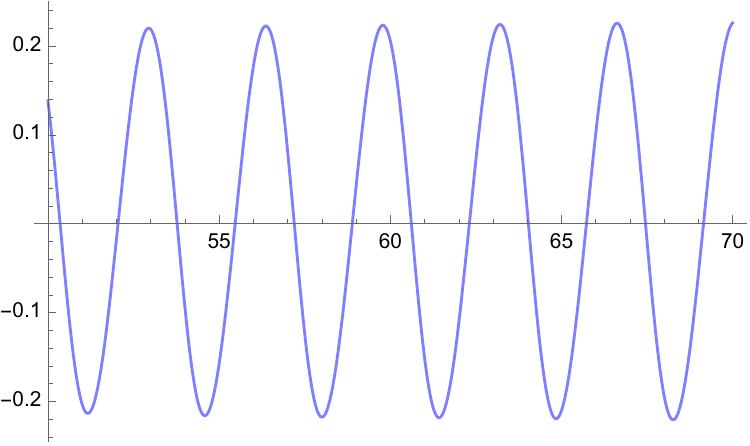} \quad
\includegraphics[width=0.2\textwidth]{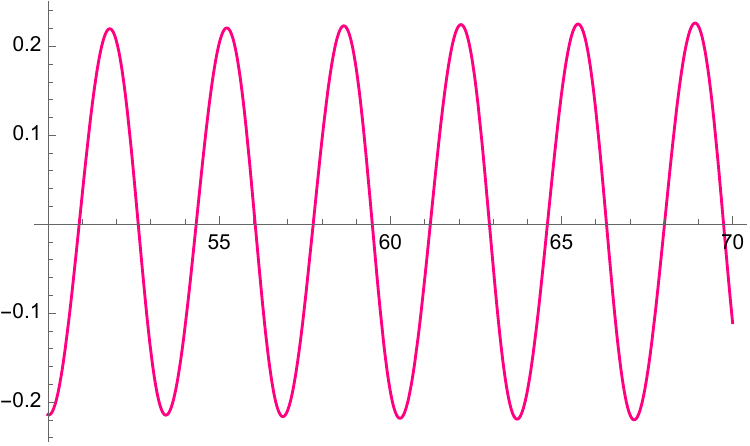} \quad
\includegraphics[width=0.2\textwidth]{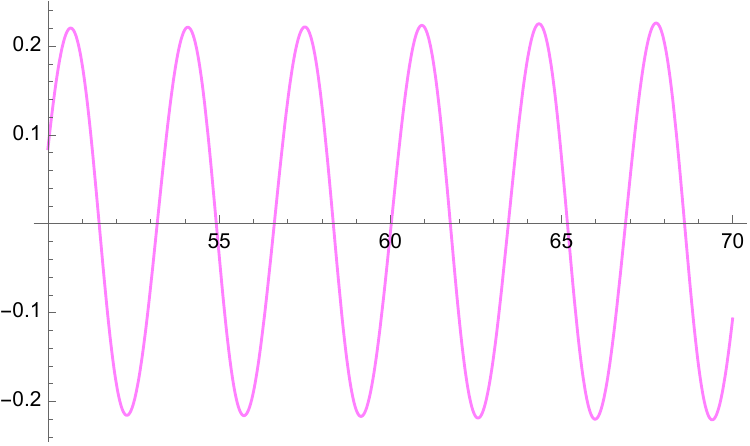} \quad
\includegraphics[width=0.2\textwidth]{FIGURES/7chainFHNZ3ALL.pdf} \quad
}
\caption{$\Z_3$ traveling wave periodic state. Time range = $[50,70]$ to eliminate transients.
{\em Top left to bottom right}: Nodes $1$--$7$; all nodes superposed.}
\label{F:7chainFHNZ3}
\end{figure}

Figure~\ref{F:7chainFHNZ3} shows a $\Z_3$ traveling wave periodic state.
Parameters are:
\begin{equation}
\label{E:FHNparams2}
I=0 \quad a =0.05 \quad b = 2.5 \quad \gamma=0.3 \quad c = -0.6
\end{equation}

\subsection{Transverse Eigenvalues}
\label{S:TE}

\begin{figure}[h!]
\centerline{%
\includegraphics[width=0.4\textwidth]{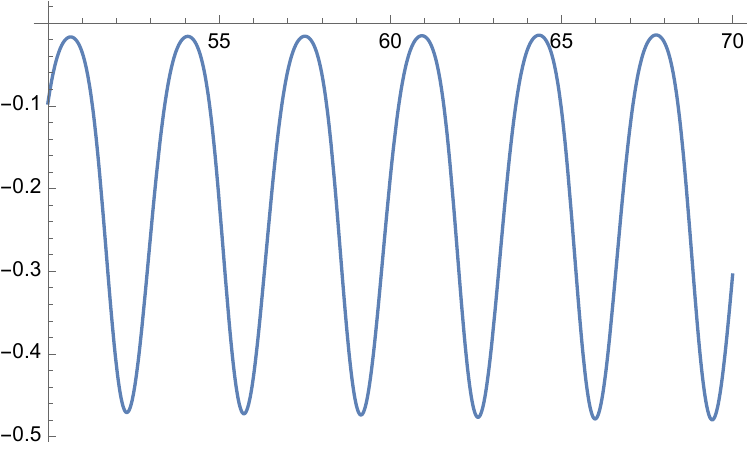} \qquad
\includegraphics[width=0.4\textwidth]{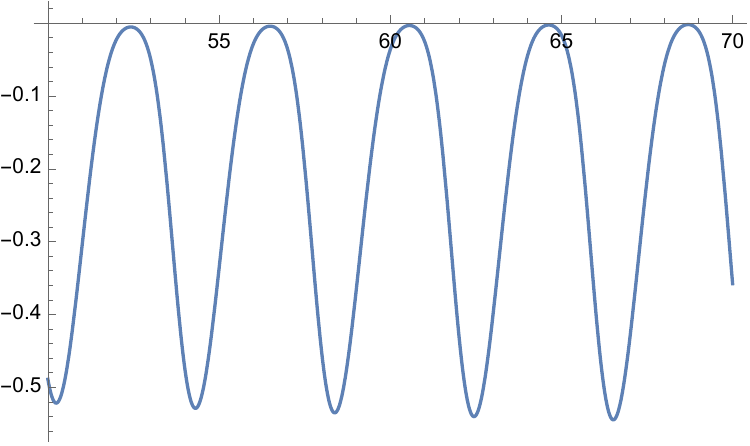}
}
\caption{
Cases where real parts of eigenvalues of $J$ at times $t \in [50,70]$ are always negative. 
Both eigenvalues have
the same real parts
because the eigenvalues are complex conjugates. {\em Left}: Parameters \eqref{E:FHNparams1}. {\em Right}: Parameters \eqref{E:FHNparams2}.}
\label{F:eigen_sync}
\end{figure}

We computed the transverse eigenvalues along the periodic orbit for the above
parameter values. The results are shown graphically in Figures
\ref{F:eigen_sync} and \ref{F:eigen_pos}, and confirm that the 
eigenvalues of the Jacobian $J$ on the periodic cycle all have negative real part, 
indicating transverse stability.
In Section \ref{S:CTFM}
we show (numerically) that these periodic orbits are transversely Floquet stable.

\begin{remark}\em
\label{r:TEneg}
If instead we use parameters
\begin{equation}
\label{E:FHNparams3}
I=0 \quad a =0.05 \quad b = 2.5 \quad \gamma=0.3 \quad c = -0.8
\end{equation}
then numerical simulations show a traveling wave very similar to 
Figure~\ref{F:7chainFHNZ3}, which therefore appears to be stable.
However, numerical calculations, Figure~\ref{F:eigen_pos} (left), show that the transverse eigenvalues
have (very small) {\em positive} real part for a short time interval.
This suggests that in practice the periodic cycle can be transversely Floquet stable
if the region of the cycle in which the transverse dynamic fails to be attractive  
is sufficiently small. Numerical calculations summarized below in Table \ref{T:FloqMultFHN}
 confirm that the CPG periodic orbit is
Floquet stable and the lifted periodic orbit is transversely Floquet stable.

\begin{figure}[h!]
\centerline{%
\includegraphics[width=0.4\textwidth]{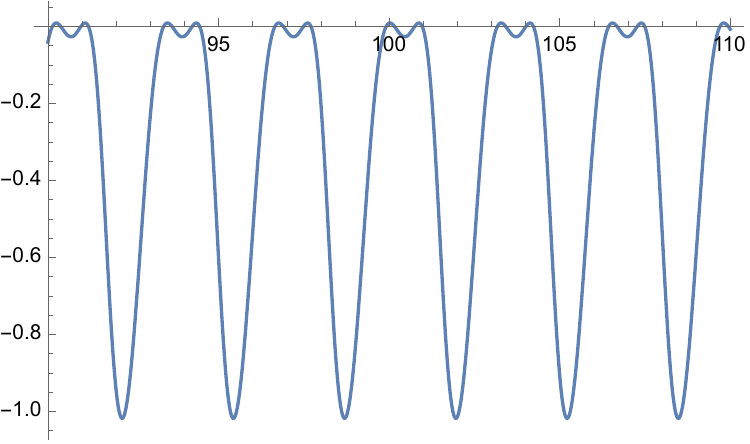} \qquad
\includegraphics[width=0.4\textwidth]{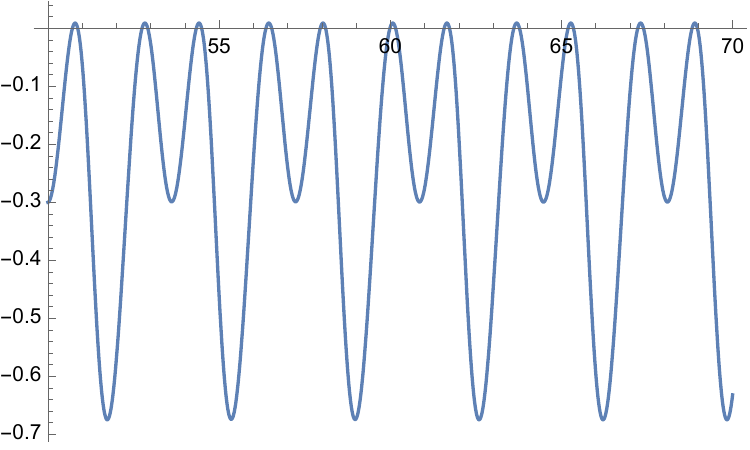} 
}
\caption{For Floquet stable periodic states,
real parts of eigenvalues of $J$ at times $t \in [50,70]$ can sometimes be positive;
see tips of double peaks. 
Both eigenvalues have
the same real parts
because the eigenvalues are complex conjugates. {\em Left}: Parameters \eqref{E:FHNparams3}. {\em Right}: Parameters \eqref{E:FHNparams4}.}
\label{F:eigen_pos}
\end{figure}

Similar results apply if the input $I$ is nonzero. Negative $I$ seems to
damp down the oscillations. Positive (not too large) $I$ leads to $\Z_3$ traveling waves,
which are apparently stable,
but again there can be a short interval of times for which the transverse eigenvalues
have (very small) positive real part. Figure~\ref{F:eigen_pos} (right) shows this for
parameters
\begin{equation}
\label{E:FHNparams4}
I=2 \quad a =0.05 \quad b = 2.5 \quad \gamma=0.3 \quad c = -0.4
\end{equation}
\end{remark}

\subsection{CPG and Transverse Floquet Multipliers}
\label{S:CTFM}

We computed the Floquet multipliers for the periodic state $a(t)$
on the CPG $\GG$.
We consider the four parameter sets \eqref{E:FHNparams1} -- \eqref{E:FHNparams4}.
The results are listed in Table 1.  
The CPG periodic orbit is
Floquet stable and the lifted periodic orbit is transversely Floquet stable, for all
four parameter sets.

\begin{table}[!htb]
\label{T:FloqMultFHN}
\begin{center}
\begin{tabular}{|l|l|l|l|l|l|}
\hline
param. & $T$ & CPG multipliers  & abs. & transverse multipliers & abs. \\
\hline
\hline
\eqref{E:FHNparams1} & $4.070$ & $1$ 
	& $1$ & $0.435$ & $0.435$ \\
& & $0.812$ & $0.812$ &0.366 & $0.366$ \\
stable & & $0.222 \pm 0.168 \,\ii$ & 0.279 & &  \\
& & $0.183 \pm 0.174\,\ii$ & 0.252 & &  \\
\hline
\eqref{E:FHNparams2}& $3.420$& $1$ & $1$ & $0.290 \pm 0.388 \,\ii$ & $0.485$\\
& & $0.868$ & $0.868$ & & \\
stable& & $-0.167 \pm 0.696 \,\ii$ & $0.715$ & & \\
& & $0.0957 \pm+ 0.141\,\ii$ & $0.171$ & & \\
\hline
\eqref{E:FHNparams3}& $3.261$ & $1$ & $1$ & $0.0850 + 0.315 \,\ii$ & $0.327$\\
& & $0.507$ & $0.507$ & & \\
stable& & $-0.444 \pm 0.401 \,\ii$ & $0.598$ & & \\
& & $-0.00298 \pm 0.0820 \,\ii$ & $0.0820$ & & \\
\hline
\eqref{E:FHNparams4}&  $3.620$ & $1$ & $1$ & $0.371 + 0.163 \,\ii$ & $0.405$ \\
&  & $0.385$ & $0.385$ & & \\
stable& & $0.233 \pm 0.532 \,\ii$ & $0.581$ & & \\
& & $0.156 \pm 0.098\,\ii$ & $0.184$ & & \\
\hline
\end{tabular}
\end{center}
\caption{CPG and transverse Floquet multipliers for FitzHugh--Nagumo examples.
The CPG periodic orbit and the lifted periodic orbit are stable for all four parameter sets.
All entries stated to three significant figures after the decimal point. abs. = absolute value of
multiplier in previous column.}
\end{table}

Comparing with the transverse stability condition, as in Section \ref{S:TE},
we see that for parameters (\ref{E:FHNparams3}, \ref{E:FHNparams4}),
transverse eigenvalues can have positive real
parts on some time intervals, without loss of transverse Floquet stability
of the lifted periodic orbit.
This suggests that some form of `transverse stability on average', 
as discussed in \cite[Section 5.2]{SW23a},
might be operating, whereby short periods of repulsion away from the synchrony
subspace are compensated for by longer periods of attraction.
However, it is unclear how to make this suggestion rigorous when node spaces
have dimension greater than $1$.

\section{Morris--Lecar Model}
\label{S:MLsim}

The Morris--Lecar model of a neuron~\cite{ML81} was originally devised
to model voltage oscillations in the muscles of barnacles. It is a simplified
variant of the Hodgkin--Huxley equations, and takes the form
\begin{equation}
\label{E:MLeq1}
\begin{array}{rcl}
C\dot{V} &=& -g_{Ca}M_{ss}(V)(V-V_{Ca})-g_KW(V-V_K)-g_L(V-V_L)+I_{app} \\
\dot{W}&=& \displaystyle\frac{W_{ss}(V)-W}{T_W(V)}
\end{array}
\end{equation}
where:

$V$ is the membrane potential

$W$ is a recovery variable (almost always the normalized K$^+$-ion conductance)

$I_{app}$ is the magnitude of stimulus current

$C>0$ is a time-constant for the voltage response.

\noindent
The conductance functions are:
\begin{equation}
\label{E:MLeq2}
\begin{array}{rcl}
M_{ss}(V)&=& (1+\tanh[(V-v_1)/v_2)])/2  \\
W_{ss}(V)&=&(1+\tanh[(V-v_3)/v_4)])/2 
\end{array}
\end{equation}
and the time constant for the K$^+$-channel relaxation is
\begin{equation}
\label{E:MLeq3}
T_W(V)=T_0\ \mathrm{sech}[(V-v_3)/2v_4]
\end{equation}
where $v_1,v_2,v_3,v_4$ are constants and
$T_0$ determines  the time scale for the recovery process.
(The $v_i$ are usually denoted by $V_1,V_2,V_3,V_4$, but
this conflicts with our notation for admissible ODEs.)

Again we simulate the 7-node network of Figure~\ref{F:7nodeFFZ3}, with a 3-node
$\Z_3$-symmetric CPG $\{1,2,3\}$ feeding forward to a chain of four nodes
$\{4,5,6,7\}$. Analytic results are not tractable for
equations (\ref{E:MLeq1},\,\ref{E:MLeq2},\,\ref{E:MLeq3}), so we exhibit
two representative numerical plots of oscillatory states with $\sot$-period phase
patterns.

For each node $c \in \{1, \ldots, 7\}$ let $P_c = \R^2$ with node coordinates
$(V_c, W_c)$.
Assume that each node obeys the Morris--Lecar equations with identical
parameters (to make the coloring balanced), and couple them via 
the voltage variables $V_c$. Then the lifted equations have the form
\beqn
\dot{V_c} &=& [-g_{Ca}M_{ss}(V_c)(V_c-V_{Ca})-g_KW_c(V_c-V_K)-g_L(V_c-V_L)+I_{app}]/C + aV_{I(c)}\\
\dot{W_c}&=& \displaystyle\frac{W_{ss}(V_c)-W_c}{T_W(V_c)}
\eeqn
Here $V_c$ is the voltage for node $c$ and $W_c$ is the recovery variable
for node $c$. The input nodes $I(c)$ are defined by
$I(1)=3$ and $I(c) = c-1$ for $2 \leq c \leq 7$.
(The notation $I(c)$ should not to be confused with $I_{app}$.)
The parameter $a$ is a coupling constant,
which we make negative to break synchrony and
obtain the pattern of  $\sot$-period phase shifts. 
Positive $a$ tends to create synchronous states, which we do not illustrate.

\subsection{Simulations}

\begin{figure}[h!]
\centerline{%
\includegraphics[width=0.4\textwidth]{FIGURES/MLnewV.pdf} \qquad\
\includegraphics[width=0.4\textwidth]{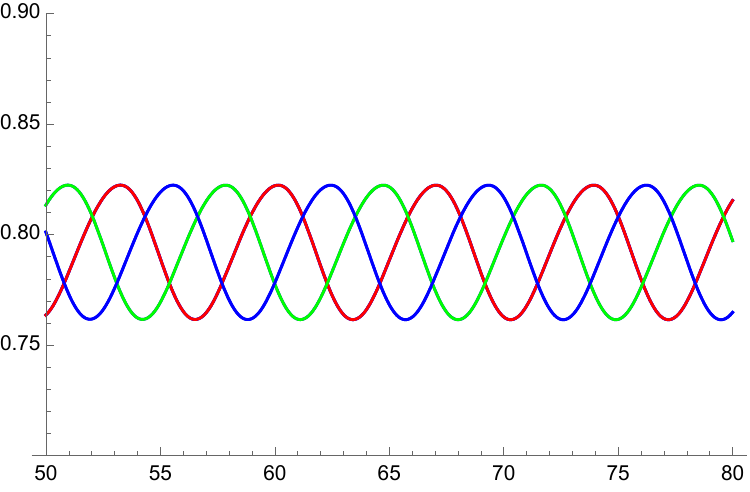}}
\caption{Morris--Lecar model. {\em Left}:  Superposed time series for $V_c$ where $1 \leq c \leq 7$.
{\em Right}: Superposed time series for $W_c$ where $1 \leq c \leq 7$.
Parameter values are \eqref{E:MLparam1}.}
\label{F:MLE}
\end{figure}

Figure~\ref{F:MLE} shows two time series for the Morris--Lecar model, namely
$V_c$ and $W_c$ where $1 \leq c \leq 7$. These plots confirm the 
required phase pattern on both CPG and chain. Only three waves appear, phase-shifted by
multiples of $\sot$ of the period, because of the synchrony pattern induced
by the balanced 3-coloring. They have the same waveform but differing phases
because of spontaneous $\Z_3$ symmetry breaking.
The parameters used are similar to those appearing in some examples in \cite{ML81},
and are listed in \eqref{E:MLparam1} and \eqref{E:MLparam2}. The time interval is $t \in [50,80]$.
In this case the waveforms are approximately sinusoidal.
Figure~\ref{F:MLE} is obtained using the parameter set:
\begin{equation}
\label{E:MLparam1}
\begin{array}{ll}
g_{Ca} = 5\quad
g_K = 8\quad
g_L = 3\quad
V_{Ca} = 7\quad
V_K = -70\quad
V_L = 50\quad
v_1 = 1\quad
v_2 = 1 \\
v_3 = -10\quad
v_4 = 14.5\quad
C = 20\quad
I_{app} = 295\quad
a = -.9\quad
T_0=5
\end{array}
\end{equation}

\begin{figure}[h!]
\centerline{%
\includegraphics[width=0.4\textwidth]{FIGURES/MLEav1.pdf} \qquad\
\includegraphics[width=0.4\textwidth]{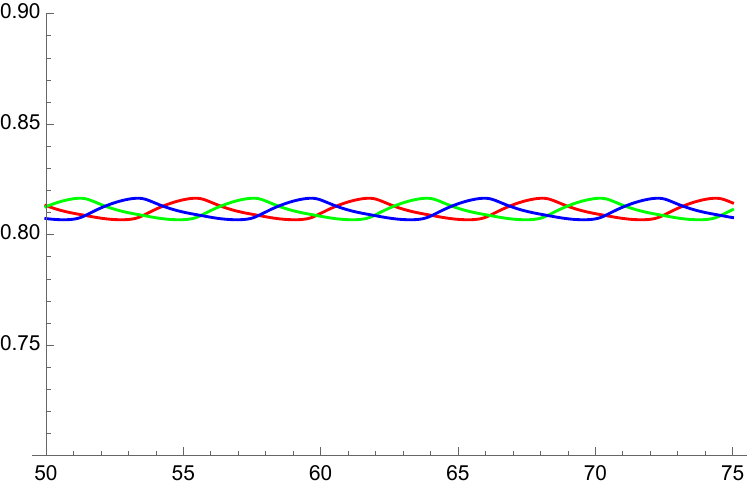}}
\caption{Morris--Lecar model. {\em Left}:  Superposed time series for $V_c$ where $1 \leq c \leq 7$.
{\em Right}: Superposed time series for $W_c$ where $1 \leq c \leq 7$.
Parameter values are \eqref{E:MLparam2}.}
\label{F:MLEav}
\end{figure}

In contrast, Figure~\ref{F:MLEav} is obtained using:
\begin{equation}
\label{E:MLparam2}
\begin{array}{ll}
g_{Ca} = 5\quad
g_K = 8\quad,
g_L = 3\quad
V_{Ca} = 7\quad
V_K = -70\quad
V_L = 50\quad
v_1 = 1\quad
v_2 = 1\\
v_3 = -10\quad
v_4 = 14.5\quad
C = 1\quad
I_{app} = 300\quad
a = -.9\quad
T_0=5
\end{array}
\end{equation}
The time interval is now $t \in [50,75]$ and vertical scales are different for clarity.
The waveforms are no longer approximately sinusoidal, but the same phase pattern
appears. 

\subsection{Transverse Eigenvalues}

\begin{figure}[h!]
\centerline{%
\includegraphics[width=0.4\textwidth]{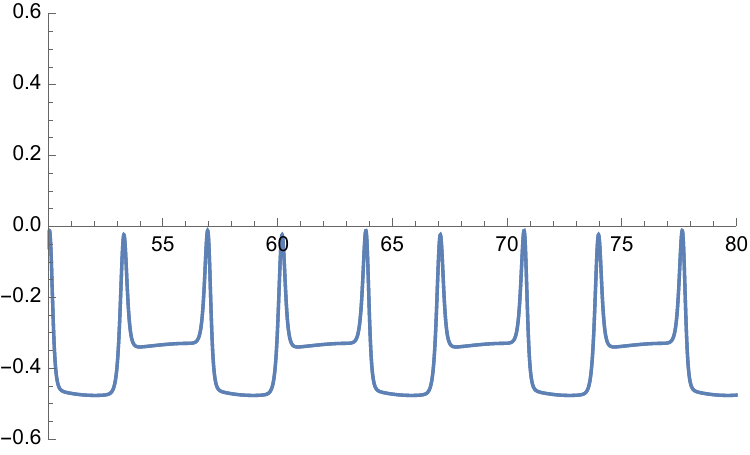} \qquad\
\includegraphics[width=0.4\textwidth]{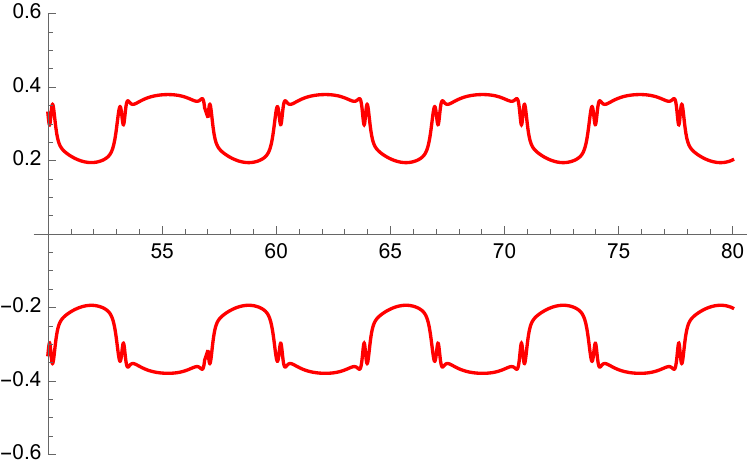}}
\caption{Morris--Lecar model. 
{\em Left}: Real parts of eigenvalues of $\Dr_{x_1}f$ on the periodic orbit $x(t)$.
{\em Right}:  Imaginary parts of eigenvalues of $\Dr_{x_1}f$ on the periodic orbit $x(t)$.
Parameter values are \eqref{E:MLparam1}.}
\label{F:MLE_TE}
\end{figure}

Figure~\ref{F:MLE_TE} shows two time series for this model, namely
the real and imaginary parts of the eigenvalues of $\Dr_{x_1}f$ on 
the first node-component of the CPG periodic orbit $x(t)$. All other components
are either in synchrony with this component or are phase-shifted versions of it.
The parameters concerned are \eqref{E:MLparam1}. The time interval is $t \in [50,80]$.
In this case the waveforms are approximately sinusoidal.
Both eigenvalues of $\Dr_{x_1}f$ have negative real parts
on the entire periodic orbit, ensuring transverse stability of the synchrony subspace.

\begin{figure}[h!]
\centerline{%
\includegraphics[width=0.4\textwidth]{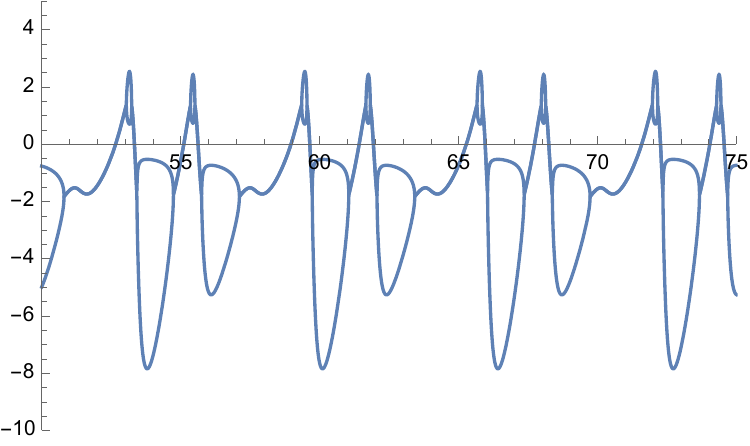} \qquad\
\includegraphics[width=0.4\textwidth]{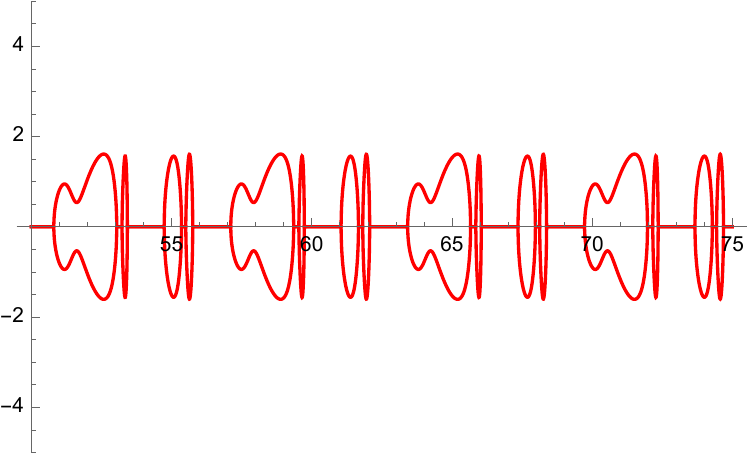}}
\caption{Morris--Lecar model.
{\em Left}: Real parts of eigenvalues of $\Dr_{x_1}f$ on the periodic orbit $x(t)$.
{\em Right}: Imaginary parts of eigenvalues of $\Dr_{x_1}f$ on the periodic orbit $x(t)$.
Parameter values are \eqref{E:MLparam2}.}
\label{F:MLEav_TE}
\end{figure}

Figure~\ref{F:MLEav_TE} shows the same time series for parameters \eqref{E:MLparam2}.
Now the real parts of some eigenvalues are positive over very short portions of the periodic orbit,
but negative on most of the orbit. This corresponds to transverse stability on average,
relative to the unique invariant probability measure on the periodic orbit.

It is also possible to find parameter values giving approximately sinusoidal
oscillations that are only transversely stable on average, so the less regular
shape of Figure~\ref{F:MLEav} (left) is not directly related to the occurence of intervals
of eigenvalues with positive real part.

\subsection{Transverse Floquet Multipliers}

The periods and transverse Floquet multipliers for the parameters above are 
listed in Table 2.
In both cases the eigenvalues lie close to the origin, 
indicating a high degree of transverse Floquet stability. This seems to be common
in the Morris--Lecar model.

\begin{table}[!htb]
\label{T:TFM_ML}
\begin{center}
\begin{tabular}{|l|l|l|l|}
\hline
parameters & period & transverse Floquet multipliers & absolute value\\
\hline
\hline
\eqref{E:MLparam1} & 6.892 & $-0.609 \pm 0.0575 \ii$ & 0.0838 \\
\hline
\eqref{E:MLparam2} & 6.309 & 0.0986 & 0.0986 \\
 & & $2.71 \times 10^{-8}$ & $2.71 \times 10^{-8}$ \\
\hline
\end{tabular}
\caption{Periods and transverse Floquet multipliers for Morris--Lecar model.
All entries stated to three significant figures after the decimal point.} 
\end{center}
\end{table}

\section{Hodgkin--Huxley Model}
\label{S:HHsim}

The Hodgkin--Huxley model of a neuron~\cite{HH52} takes the form
\beqn
C_m \dot{V_m}&=& I - [ \bar{g}_K n^4 ( V_m - V_K ) + \bar{g}_{Na} m^3 h (V_m - V_{Na}) + \bar{g}_l (V_m - V_l)] \\   
\dot{n} &=& \alpha_n ( V_m ) ( 1 - n ) - \beta_n ( V_m ) n  \\
\dot{m} &=&\alpha_m ( V_m ) ( 1 -m ) - \beta_m ( V_m ) m \\
\dot{h} &=& \alpha_h ( V_m ) ( 1 - h ) - \beta_h ( V_m ) h     
\eeqn
Here the variables are:

\quad $V_m$ is the membrane potential (voltage).

\quad $n, m$, and $h$ are dimensionless quantities between 0 and 1
associated with potassium channel activation, sodium channel activation, 
and sodium channel inactivation, respectively. 
There are also numerous parameters:

\quad $C_m$ is the membrane capacitance.

\quad $I$ is the current per unit area.
 
\quad $\alpha _{i}$ and $\beta _{i}$ are rate functions for the $i$th ion channel.

\quad $\bar {g}_{n}$ is the maximal value of the conductance.

In the original paper~\cite{HH52},
$\alpha _{i}$ and $\beta _{i}$ (for $i = n , m , h$) are given by:    
\beqn
\alpha _{n}(V_{m})&=& {\frac {0.01(V_{m}+10)}{\exp {\big (}{\frac {V_{m}+10}{10}}{\big )}-1}}\\
\alpha _{m}(V_{m})&=&{\frac {0.1(V_{m}+25)}{\exp {\big (}{\frac {V_{m}+25}{10}}{\big )}-1}}\\
\alpha _{h}(V_{m})&=&0.07\exp {\bigg (}{\frac {V_{m}}{20}}{\bigg )}\\
\beta _{n}(V_{m})&=&0.125\exp {\bigg (}{\frac {V_{m}}{80}}{\bigg )} \\
\beta _{m}(V_{m})&=&4\exp {\bigg (}{\frac {V_{m}}{18}}{\bigg )} \\
\beta _{h}(V_{m})&=&{\frac {1}{\exp {\big (}{\frac {V_{m}+30}{10}}{\big )}+1}}
\eeqn

\subsection{Simulations}

\begin{figure}[h!]
\centerline{%
\includegraphics[width=0.4\textwidth]{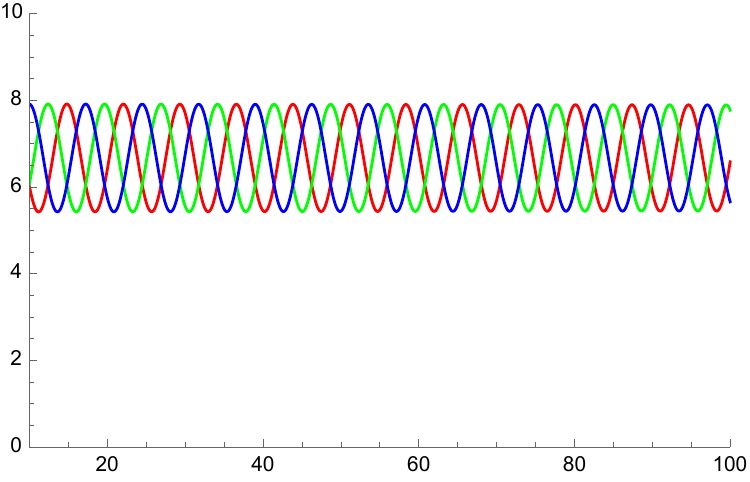} \qquad\
\includegraphics[width=0.4\textwidth]{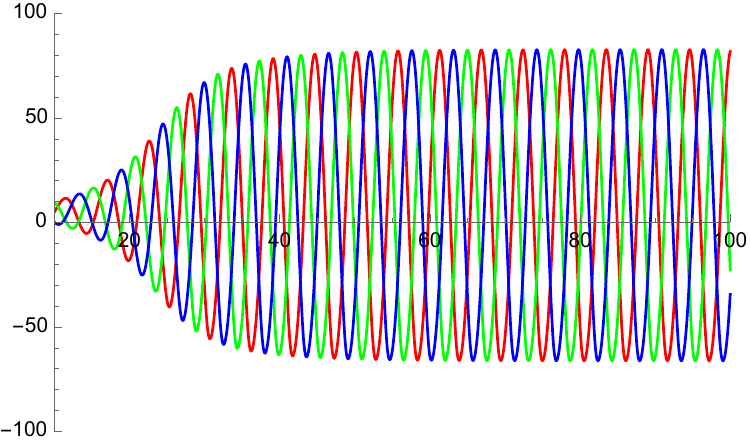}
}
\caption{Hodgkin--Huxley. Superposed time series for $V_c$ where $1 \leq c \leq 7$.
{\em Left}: Parameter values as in \eqref{E:HHparam1}. {\em Right}: Parameter values as in \eqref{E:HHparam2}; initial transients are included to illustrate convergence onto the
periodic  state.
}
\label{F:HHplots}
\end{figure}

We simulated the 7-node network for the Hodgkin--Huxley model, 
where each node is a Hodgkin--Huxley neuron
with identical parameters, voltage-coupled with coupling constant $a$.
Again we make $a$ negative to break synchrony and
obtain the pattern of  $\sot$-period phase shifts. The parameters we used are:

\begin{equation}
\label{E:HHparam1}
\begin{array}{l}
\bar{g}_K = 10 \quad
\bar{g}_{Na} = 20\quad
\bar{g}_l = 20\quad 
V_K = -150\quad
V_{Na}=100\quad \\
V_l = 20\quad
C_m = 40\quad
I = 0\quad
a = -1
\end{array}
\end{equation}

\begin{equation}
\label{E:HHparam2}
\begin{array}{l}
\bar{g}_K = 100 \quad
\bar{g}_{Na} = 50\quad
\bar{g}_l = 20\quad
V_K = 50\quad
V_{Na}=10\quad\\
V_l = 20\quad 
C_m = 40\quad
I = -50\quad
a = -1.3
\end{array}
\end{equation}

 \noindent
Figure~\ref{F:HHplots} shows superposed time series for this model, for the voltages
$V_c$ where $1 \leq c \leq 7$, all superposed. As usual we see only three traces,
with $\sot$-period phase shifts, confirming the required phase pattern.

\subsection{Transverse Eigenvalues}
  
\begin{figure}[h!]
\centerline{%
\includegraphics[width=0.4\textwidth]{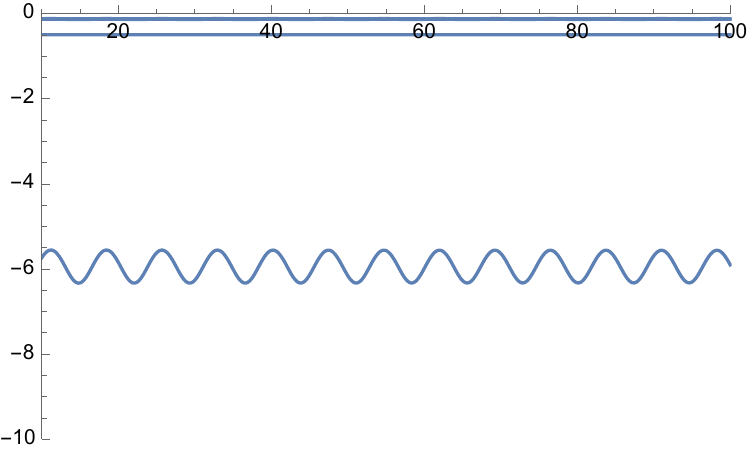} \qquad\ 
\includegraphics[width=0.4\textwidth]{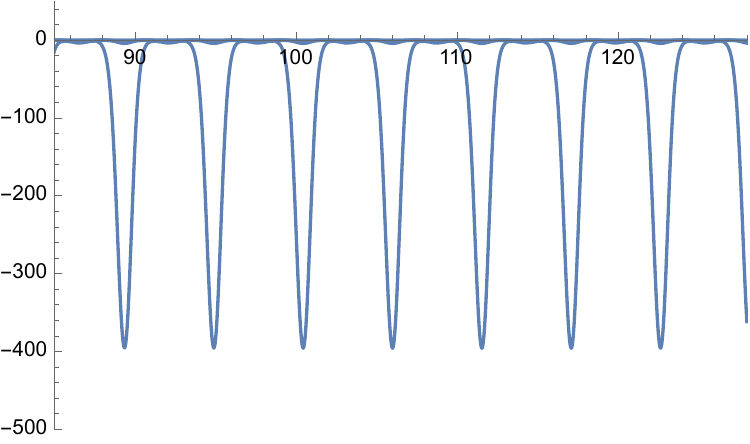} 
}
\caption{Hodgkin--Huxley. 
Real parts of eigenvalues of $\Dr_{x_1}f$ on the periodic orbit $x(t)$.
{\em Left}: Parameter values as in \eqref{E:HHparam1}. {\em Right}: Parameter values as in \eqref{E:HHparam2}.
}
\label{F:HHreplots}
\end{figure}

Figure~\ref{F:HHreplots} shows the corresponding time series for the
real parts of eigenvalues of $\Dr_{x_1}f$ on the periodic orbit $x(t)$. 
(Here $x(t)$ can be either
the CPG orbit or the lift, since all seven nodes are phase-synchronous. For the
same reason we do not need to compute eigenvalues of $\Dr_{x_c}f$ for
$c > 1$.)
The figure appears to show only three eigenvalues in each case, 
but inspection of the numerical values shows that the top line
actually consists of two lines very close together representing two eigenvalues
that are almost the same, but not exactly so. All four eigenvalues have
negative real part on the entire orbit, implying transverse stability of the synchrony
subspace.

\begin{figure}[h!]
\centerline{%
\includegraphics[width=0.4\textwidth]{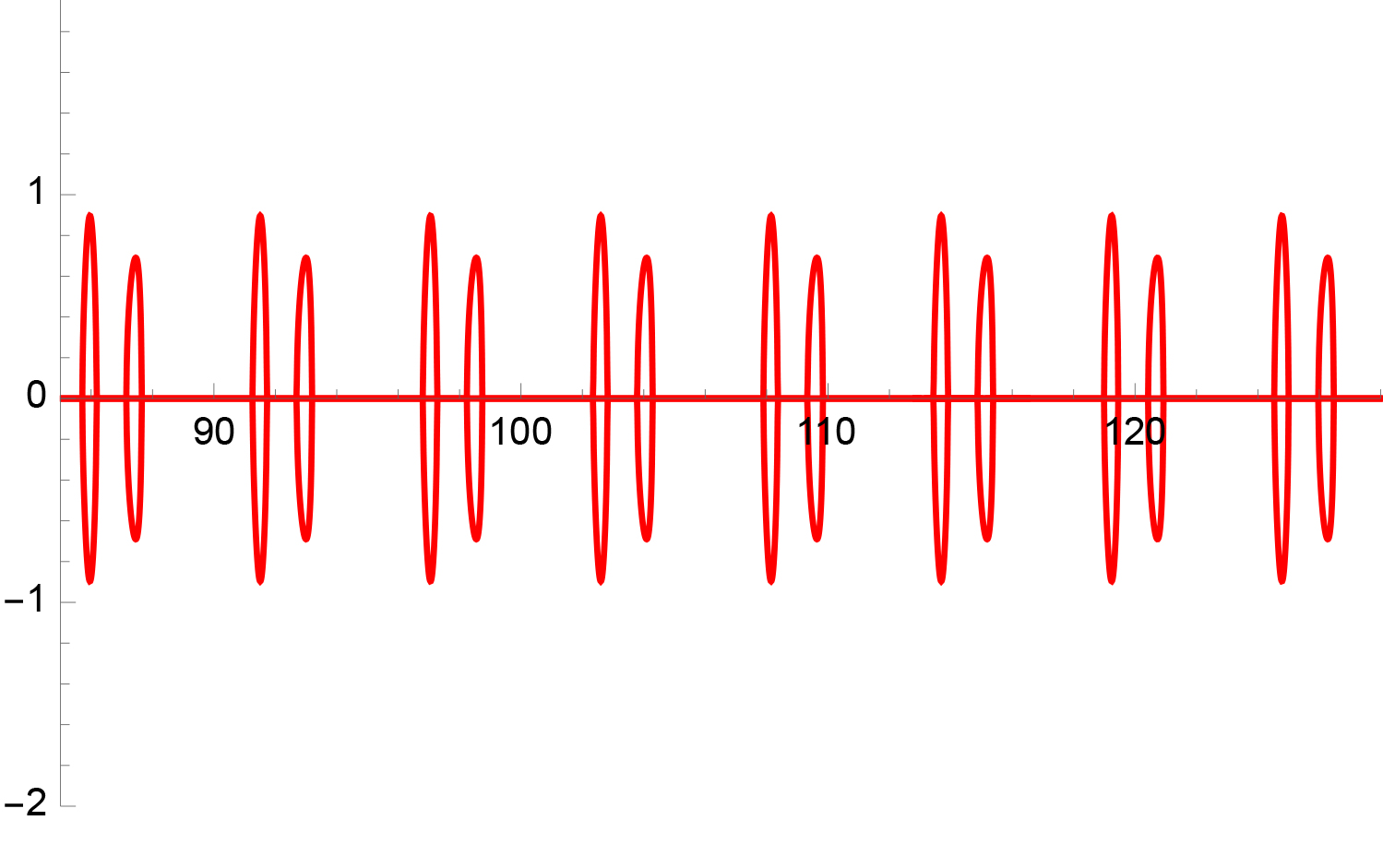} \qquad\ 
\includegraphics[width=0.4\textwidth]{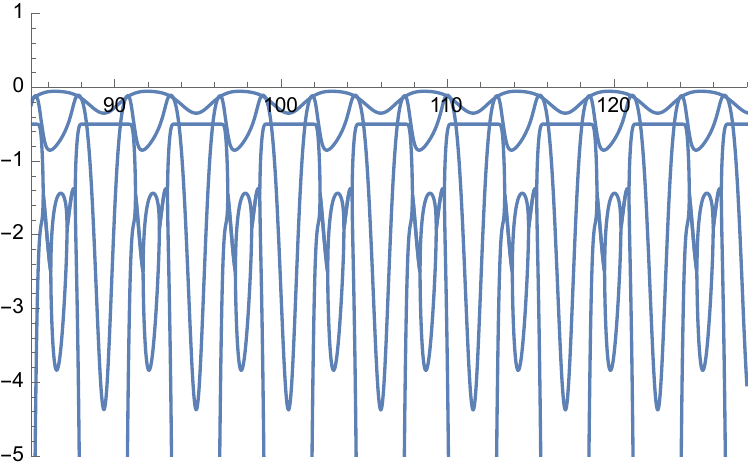} 
}
\caption{Hodgkin--Huxley.  Parameter values as in \eqref{E:HHparam2}.
{\em Left}: Imaginary parts of eigenvalues of $\Dr_{x_1}f$ on the periodic orbit $x(t)$.
 {\em Right}: Larger scale time series for real
  parts of eigenvalues of $\Dr_{x_1}f$ on the periodic orbit $x(t)$. 
}
\label{F:HHimplots}
\end{figure}

For parameters \eqref{E:HHparam1},
the imaginary parts of these eigenvalues are all zero, so are not plotted.
For parameters \eqref{E:HHparam2}, the imaginary parts of these eigenvalues are not all zero,
and are plotted in Figure \ref{F:HHimplots} (left). The real parts, Figure \ref{F:HHimplots} (right),
are all negative.

We have not found   
 parameters giving a periodic orbit of the Hodgkin--Huxley model
 with only transverse stability on average. The transverse eigenvalues
have negative real parts in all cases we have computed.

\subsection{Transverse Floquet Multipliers}

Table 3 shows computed transverse Floquet multipliers
for the Hodgkin--Huxley model with parameters \eqref{E:HHparam1} and \eqref{E:HHparam2}.
Both cases are transversely Floquet stable.

\begin{table}[!htb]
\label{T:TFM_HH}
\begin{center}
\begin{tabular}{|l|l|l|l|}
\hline
parameters & period & transverse Floquet multipliers & absolute value\\
\hline
\hline
\eqref{E:HHparam1} & 7.253 & $0.410$ & 0.410 \\
	& & 0.373& 0.373\\
	& & 0.0265 & 0.0265 \\
	& & $-8.03 \times 10^{-11}$ & $8.03 \times 10^{-11}$\\
\hline
\eqref{E:HHparam2} & 5.546 & $0.387$ &0.387 \\
	& & 0.0941& 0.0941 \\
	& & $1.88\times 10^{-6}$ & $1.88\times 10^{-6}$ \\
	& & $2.47 \times 10^{-8}$ &  $2.47 \times 10^{-8}$ \\
\hline
\end{tabular}
\caption{Periods and transverse Floquet Multipliers for Hodgkin--Huxley model.
All entries stated to three significant figures after the decimal point.}
\end{center}
\end{table}

\section{Hindmarsh--Rose Model}
\label{S:HRM}

The Hindmarsh--Rose neuron model was originally introduced in \cite{HR84}
as a simplification of the Hodgkin--Huxley equations, with the main focus
on spiking activity. The equations have the form
\begin{equation}
\label{E:HReq}
\begin{array}{rcl}
\dot{x} &=& y -ax^3+bx^2-z+I \\
\dot y &=& c-dx^2-y \\
\dot z &=& r[s(x-x_R)-z]
\end{array}
\end{equation}
Here 

\quad $x$ is the membrane potential.

\quad $y,z$ are variables modelling ion channel dynamics.

\noindent
All other quantities are parameters:

\quad $a,b,c,d$ model fast ion channels.

\quad $r$ models slow ion channels.

\quad $I$ is the input current to the neuron.

\quad $x_R$ sets an equilibrium point.

\noindent
The Hindmarsh--Rose equations can generate a variety of dynamic behaviour,
including chaos.
Standard parameters are $s=4, x_R=-1.6, a=1,b=3,c=1,d=5$.
The parameter $r \sim 10^{-3}$ and $-10 \leq I \leq 10$.

\subsection{Simulations}

\begin{figure}[h!]
\centerline{%
\includegraphics[width=0.4\textwidth]{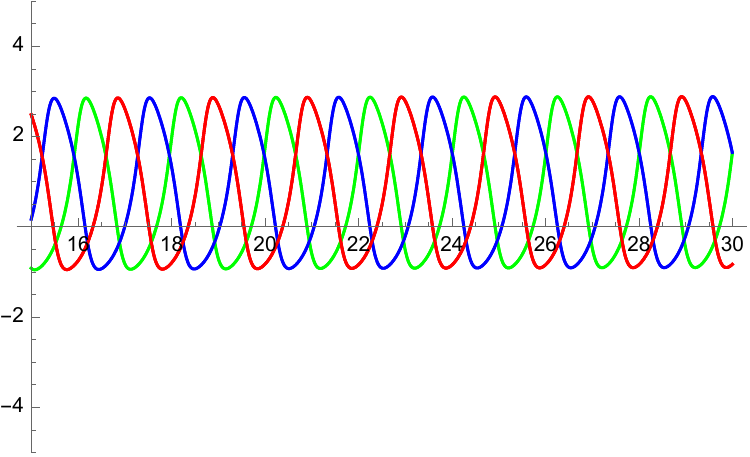}\qquad\
\includegraphics[width=0.4\textwidth]{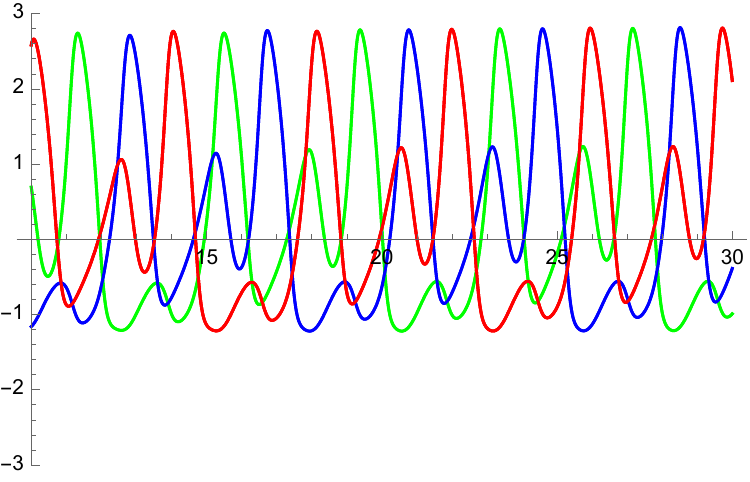}
}
\vspace{.1in}
\centerline{%
\includegraphics[width=0.4\textwidth]{FIGURES/HRts3.pdf}
}
\caption{Three simulations of the 7-node feedforward chain of Figure~\ref{F:7nodeFFZ3}
with Hindmarsh--Rose dynamics \eqref{E:HReq} on each node. Parameters are:
{\em Top left}: \eqref{E:HRparam1}.
{\em Top right}: \eqref{E:HRparam2}.
{\em Bottom}: \eqref{E:HRparam3}.
}
\label{F:HR}
\end{figure}

Figure~\ref{F:HR} shows three simulations of the 7-node chain. We use
Hindmarsh--Rose dynamics \eqref{E:HReq} on each node $c$ with variables $(x_c,y_c,z_c)$.
Coupling is by adding a term $g x_{i(c)}$ to the $x_c$-component of each
node, where $i(c)$ is the (unique) node inputting to $c$.

The figure is
a superposition of the $x$-coordinates of nodes 1--7, verifying the lifted
property of the $\sot$-period phase pattern. 
The respective parameter values are:
\begin{eqnarray}
\label{E:HRparam1}
&&a = 1 \  b = 3\  c = 1\  d = 5\  r = 0.1\  s = 0\  x_R = -1.6\  I = 10
\  g = -1 \\
\label{E:HRparam2}
&&a = 1\  b = 3\  c = 1\  d = 5\  r = 0.1\  s = 0 \  x_R = -1.6\  I = 5 \  g = -2 \\
\label{E:HRparam3}
&&a = 1\  b = 4\  c = 2\  d = 5\  r = 0.05\  s = 2\   x_R = -1.6\  I = 5\  g = -2
\end{eqnarray}
The top figure shows roughly sinusoidal waveforms. The middle figure
shows more complex waveforms. Between each peak of the time series for
a single node two different waveforms alternate---possibly some kind of period-doubling effect. The bottom figure shows spiking waveforms, which was the original aim of the
Hindmarsh--Rose model.

\subsection{Transverse Eigenvalues}

\begin{figure}[h!]
\centerline{%
\includegraphics[width=0.4\textwidth]{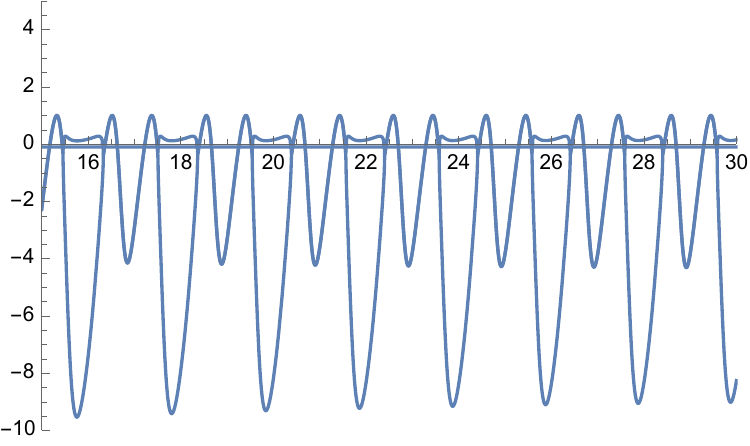} \qquad\ 
\includegraphics[width=0.4\textwidth]{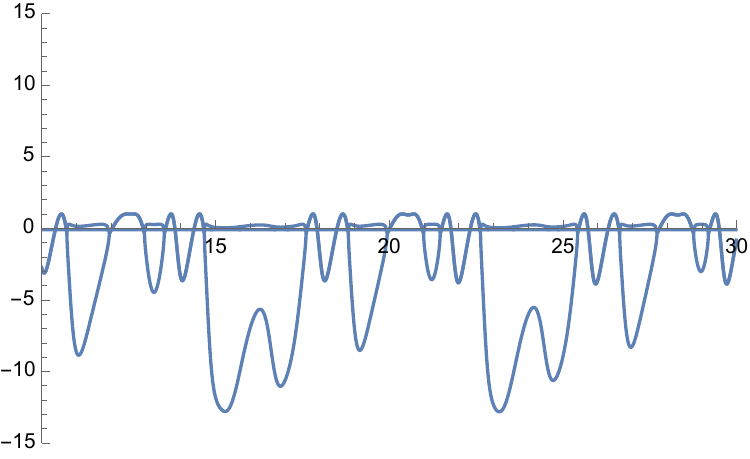}
}
\vspace{.1in}
\centerline{%

\includegraphics[width=0.4\textwidth]{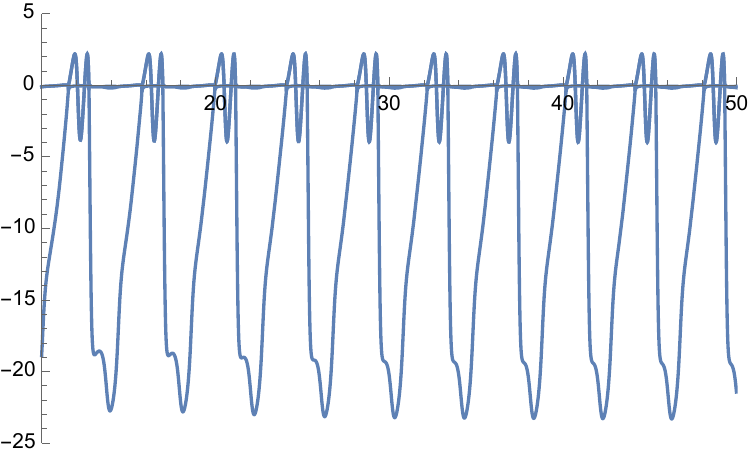}
}
\caption{Transverse eigenvalues for the 7-node feedforward chain of Figure~\ref{F:7nodeFFZ3}
with Hindmarsh--Rose dynamics \eqref{E:HReq} on each node. Parameters are:
{\em Top left}:  \eqref{E:HRparam1}.
{\em Top right}: \eqref{E:HRparam2}.
{\em Bottom}: \eqref{E:HRparam3}.
}
\label{F:HRev}
\end{figure}

Figure \ref{F:HRev} plots the real parts of the three transverse eigenvalues.
In all three cases there are short intervals over which the real part of at least one
eigenvalue is positive, and shorter intervals over which all real parts of
eigenvalues are negative. However, the real parts of the
negative eigenvalues are much larger on those intervals. Presumably this
leads to transverse stability on average, in some suitable sense.

\subsection{Transverse Floquet Multipliers}

Table 4 shows computed transverse Floquet multipliers
for the Hindmarsh--Rose model with parameters \eqref{E:HRparam1},
\eqref{E:HRparam2}, and \eqref{E:HRparam3}. All three cases are
transversely Floquet stable.

\begin{table}[h!]
\begin{center}
\label{T:TFM_HR}
\begin{tabular}{|l|l|l|l|}
\hline
parameters & period & transverse Floquet multipliers & absolute value\\
\hline
\hline
\eqref{E:HRparam1} & 1.973 & $0.820$ & 0.820 \\
	& & 0.187 & 0.187 \\
	& & 0.00465 & 0.00465 \\
\hline
\eqref{E:HRparam2} &3.554 & 0.700 & 0.700 \\
 & & 0.270 & 0.270 \\
 & & $2.78 \times 10^{-6}$ & $2.78 \times 10^{-6}$
 \\
\hline
\eqref{E:HRparam3} & 3.919 & 0.887 & 0.0887 \\
 & &  0.534 & 0.534 \\
 & & $4.40 \times 10^{-9}$ & $4.40 \times 10^{-9}$ \\
\hline
\end{tabular}
\caption{Periods and transverse Floquet Multipliers for Hindmarsh--Rose model.
All entries stated to three significant figures after the decimal point.}

\end{center}
\end{table}

\section{Breaking Synchrony}
\label{S:BS}

The assumption that some nodes and arrows in the network have the same types,
leading to a balanced coloring in which nodes are {\em exactly} synchronized, 
is an idealization. In real networks, especially biological ones, nodes and arrows 
may be very similar, but not identical, and nodes are considered synchronous if
they exhibit very similar---but not identical---time series. Formalizing
suitable notions of `approximate symmetry' and `approximate synchrony'
is possible (such as the `pseudosymmetries' of \cite{MM19}), but lacks the elegant mathematical structure of the exact
versions.

The term `symmetry breaking' is used in two distinct ways \cite{GS83}.
In {\em spontaneous} symmetry-breaking the symmetry of the model equations
is exact but solutions have less symmetry because the fully symmetric state becomes unstable.
In {\em forced} (or {\em induced}) symmetry-breaking the symmetry of the model equations
is only approximate. A similar distinction applies to `synchrony-breaking'.
Transverse (Floquet) stability prevents spontaneous synchrony-breaking in
an idealized feedforward chain. We now study forced synchrony-breaking.

We prefer to consider the idealized models as useful 
approximations, whose solutions, suitably interpreted, offer insights in to the
more messy reality. If the ideal model has exactly synchronous nodes,
small changes to the model that break synchrony generally lead to
approximately synchronous nodes. By the Kupka--Smale Theorem \cite{K64,S63,P66}, 
steady or periodic states of a dynamical system 
are generically hyperbolic: see \cite{GH83} for equilibria and
\cite{KH95} for equilibria and periodic states. All Floquet stable periodic orbits are hyperbolic.
Hyperbolic states persist under sufficiently small $C^1$ perturbations \cite{HPS77}.
An equivariant version of the Kupka--Smale Theorem is proved in \cite{F77,F80}, 
which applies to
symmetric dynamical systems. It is plausible that a similar
genericity theorem applies to networks, but this has been 
proved only for a few small networks \cite{S22}.

These theoretical considerations are supported by numerical experiments,
in which idealized equations are modified to break symmetry and synchrony.
These experiments suggest that in
practice the traveling wave states obtained from balanced colorings
are quite robust, and similar states
persist if the nodes and couplings differ---though not by too large an amount.
Exact synchrony or phase relations are replaced by approximate ones.
There are many variations with similar properties. In the next subsections we 
expand on these theoretical considerations and report results for
some typical examples of such experiments.

\subsection{Theoretical Considerations}

To keep the discussion within bounds we consider only Floquet stability.
One important feature of a feedforward lift is that it propagates
periodic CPG signals exactly, and that transverse stability one step along the chain
implies transverse stability along the entire chain. When combined with stability
of the CPG signal, this guarantees that the entire propagating signal is stable.
Another important feature is that regular phase patterns in the CPG
generate the same phase patterns in the chain, so that discrete rotating waves
in the CPG generate apparent traveling waves along the chain.

These results are proved above only for the idealized feedforward chain,
where all nodes and couplings in the chain are identical to appropriate ones
in the CPG. In a real biological system such conditions at best hold approximately.
So we can ask: what effect do such approximations have on the
existence, stability, phase patterns, and amplitudes of the propagating
signal?

These questions can be formulated in many ways. In this subsection we offer some 
theoretical reasons why we expect the results to be relatively robust,
provided exact equalities of waveform are replaced by suitable
approximate equalities. These considerations 
address only `sufficiently small' changes to dynamics and couplings.
{\em How} small is sufficient depends on the model ODE. We discuss
some typical quantitative answers, obtained by simulation, in the next
subsection. These simulations, and many others not mentioned here, indicate that
in practice the feedforward chain propagation mechanism is remarkably
robust, provided that the CPG oscillation that drives the chain continue
to exist and has a reasonably large amplitude.

When the oscillation is Floquet stable, the correponding periodic orbit
is hyperbolic. Therefore this orbit persists, remains stable, and perturbs continuously,
for any continuous family of $C^1$-small perturbations of the ODE \cite{HPS77}. Such
perturbations include:

(a)	Changes to CPG node dynamics.

(b)	Changes to CPG couplings.

(c)	Changes to node dynamics in the chain.

(d)	Changes to couplings in the chain.

\noindent
Moreover, such changes can be made independently for each node and arrow.

A small deformation of the periodic orbit projects onto node coordinates
to give small deformations of the waveforms observed at each node.
Therefore phases and amplitudes change by small amounts. In other words,
the main qualitative features of the propagating signal persist, 
with exact synchrony replaced by approximate synchrony,
and changes in quantitative features are small.

Indeed, we expect these features to persist until the perturbation becomes large enough to
cause a bifurcation. If it is a local bifurcation, which is a reasonable assumption
in many cases, it is caused by some Floquet multiplier becoming unstable.
The numerical calculations reported
in Sections \ref{S:FHNM}--\ref{S:HRM} show that the
transverse Floquet multipliers are often {\em very} stable; that is, well inside the unit circle.
For Fitzhugh--Nagumo neurons the largest (absolute value of a)
transverse eigenvalue in Table \ref{T:FloqMultFHN}
is $0.435$. For Morris--Lecar neurons the largest  transverse eigenvalue in Table \ref{T:TFM_ML} is $0.0986$. For Hodgkin--Huxley neurons the largest transverse eigenvalue in Table \ref{T:TFM_HH} is $0.410$. For Hindmarsh--Rose neurons the largest transverse eigenvalue in Table \ref{T:TFM_HR} is $0.820$. Typically the transverse eigenvalues
with smaller absolute value are much smaller. 

These figures suggest that a signal propagating along 
a chain of Hindmarsh--Rose neurons will be more fragile 
(to perturbations of the model) than
one propagating along a chain of Fitzhugh--Nagumo or Hodgkin--Huxley neurons,
while one propagating along a chain of Morris--Lecar neurons is likely to be
more robust. Possibly the largest absolute value of a
transverse eigenvalue could be used as a rule-of-thumb `stability index',
with smaller indices suggesting more robust propagation. This
would be analogous to the use of the maximal Liapunov exponent
in the `master stability function' for synchronous equilibria \cite{ADKMZ08,PC98}.

The key question in this paper is: given a stable CPG oscillation, how can it be
propagated stably along a chain, in a sufficiently robust manner
for biological and other applications? Perturbations (a) and (b) above are changes
to the CPG dynamics, and can be dealt with by recalculating the transverse Floquet
multipliers. That is, we can work with a specified CPG, and assume that it is
fixed. Attention therefore focuses on (c) and (d), changes along the chain.

The most important case is a change that affects the {\em first} module along the chain.
If the first change occurs at a later step, this is equivalent to replacing the
CPG oscillation by the signal transmitted by the previous module. Initially, it
therefore make sense to consider (c) and (d) for the first module in the chain. 
That is, node 4 in Figure \ref{F:7nodeFFZ3} and the arrow into it from node 3. 
We therefore focus on this case, which illustrates the main issue of
robustness to perturbations by terms that are not admissible and thus break synchrony.
However, we also examine some features for longer chains (up to 30 nodes).
Numerical experiments on other modifications could be continued indefinitely.

\subsection{Forced Symmetry Breaking: Fitzhugh-Nagumo}

As an exemplar we take the system of equations for the 7-node model using Fitzhugh-Nagumo equations for each node as outlined in Section \ref{S:FHNM}, Equations (\ref{E:7nodeFHN}), with the completely balanced case having parameters (\ref{E:FHNparams1}) for each node. We reproduce these equations here for convenience:
\begin{eqnarray*}
G(V,W,I) &=& V (a - V) (V - 1) - W + I \\
H(V,W) &=& bV - \gamma W
\end{eqnarray*}
with
\begin{equation*}
\begin{array}{rcl}
\dot{V}_1 &=& G(V_1,W_1,I)+cV_3 \qquad \dot{W}_1 = H(V_1,W_1) \\
\dot{V}_2 &=& G(V_2,W_2,I)+cV_1 \qquad \dot{W}_2 = H(V_2,W_2)\\
\dot{V}_3 &=& G(V_3,W_3,I)+cV_2 \qquad \dot{W}_3 = H(V_3,W_3)\\
\dot{V}_4 &=& G(V_4,W_4,I)+cV_3 \qquad \dot{W}_4 = H(V_4,W_4)\\
\dot{V}_5 &=& G(V_5,W_5,I)+cV_4 \qquad \dot{W}_5 = H(V_5,W_5)\\
\dot{V}_6 &=& G(V_6,W_6,I)+cV_5 \qquad \dot{W}_6 = H(V_6,W_6)\\
\dot{V}_7 &=& G(V_7,W_7,I)+cV_6 \qquad \dot{W}_7 = H(V_7,W_7)
\end{array}
\end{equation*}
Initially we assumed parameter values
\begin{equation*}
I=0 \quad a =0.05 \quad b = 2.5 \quad \gamma=0.3 \quad c = -0.6
\end{equation*}

These equations were then run through Matlab using its {\tt ode89} solver, an implementation of Verner's `most robust' Runge--Kutta 9(8) pair with an 8th-order continuous extension \cite{V06}.
We used random initial conditions, whilst for each run step changing one of the parameters in turn, $b, \gamma, c$ for the fourth node in the network by factors from $0.5$ up to $2$. Thus we are looking at the effect of changing both the internal dynamics of, and the coupling strength to, this node. As remarked in the previous subsection, 
 there is nothing special about this choice of node; doing the same to any node further down the chain will lead to an identical response both for that node and for nodes downstream from that one. Additional runs were made for a selection of parameter choices for much longer chains, extending the 7-node chain to up to 30 nodes to confirm how these later nodes behave with respect to the changes to node 4.

Before perturbing any of the parameters, the time series for the seven nodes were superposed, again demonstrating a traveling wave pattern with phase differences of one third of a period, see Figure \ref{F:changec} (top left).

In all cases the periods of all nodes remained almost constant to a remarkable extent, so that phase difference remained a meaningful concept even in the case of different waveforms (chiefly different maximum amplitudes). We used the times of maximum amplitude to calculate phase differences between nodes.

\subsubsection{Changing the Coupling Term $c$}

\begin{figure}[h!]
\centerline{
\includegraphics[width=0.4\textwidth]{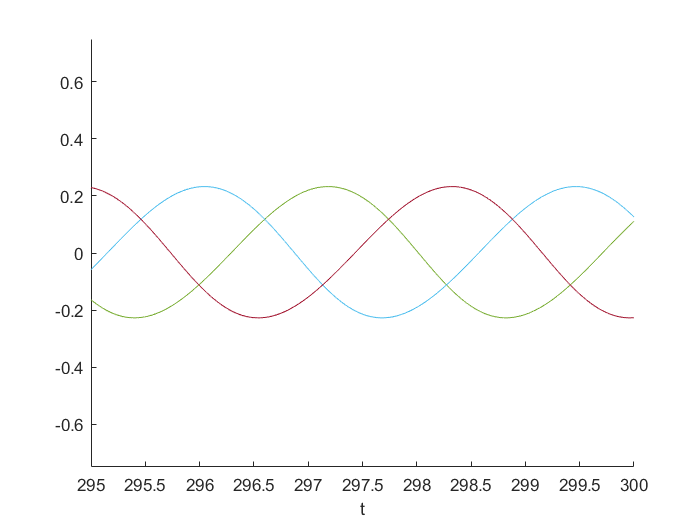} \qquad
\includegraphics[width=0.4\textwidth]{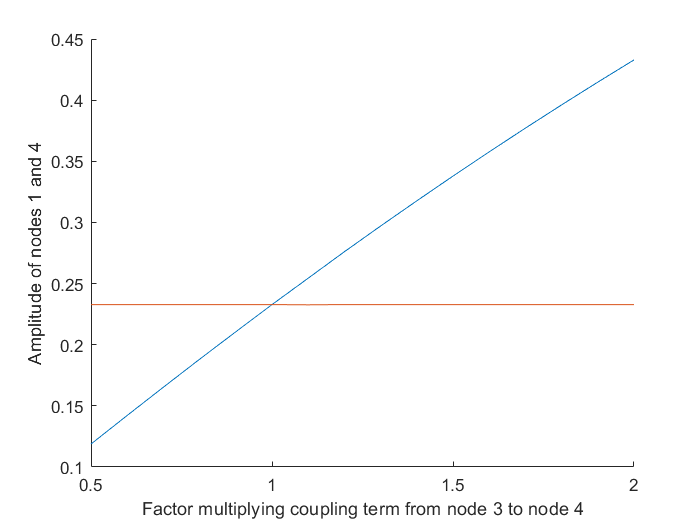}
}
\centerline{%
\includegraphics[width=0.4\textwidth]{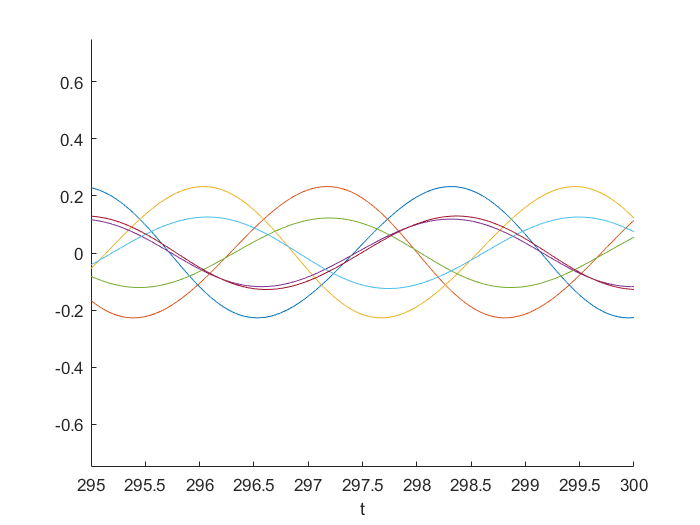} \qquad
\includegraphics[width=0.4\textwidth]{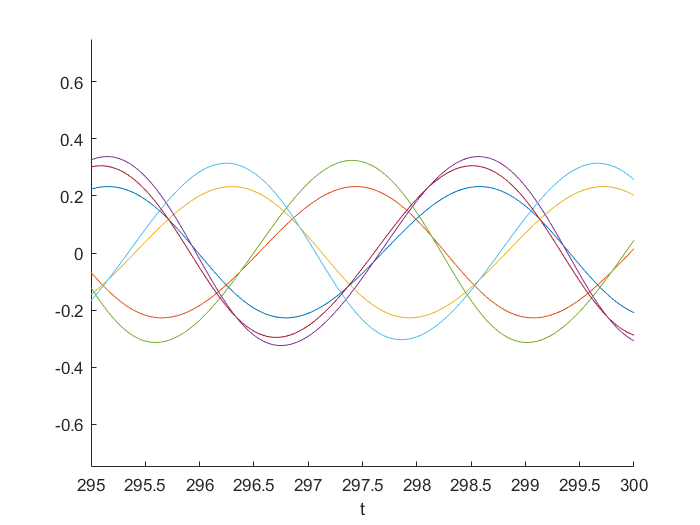}
}
\caption{Traveling wave periodic states in the 7-node FitzHugh--Nagumo feedforward network. The top subfigures show ({\em left}) all seven nodes superposed before perturbing parameters  and ({\em right}) the effect on the amplitude of the fourth node when changing the coupling from the third node (blue) compared to the amplitude of nodes 1 to 3 (red). The bottom subfigures show all 7 nodes superposed when the same coupling term is multiplied by a factor of 0.5 ({\em left}) and 2 ({\em right}).
}
\label{F:changec}
\end{figure}

The maximum amplitude for all nodes when there are no forced symmetry breaking perturbations is 0.233. Changing the coupling term from node 3 to node 4 by the above factors alters the maximum amplitude of node 4 from 0.119 to 0.433, with an approximately linear increase as the value of $c$ is increased from the lower to upper values (see Figure \ref{F:changec}, top right). This change in amplitude propagates along the chain, gradually converging towards the original values; for example, the amplitude of node 7 varies from 0.13 to 0.338. As more nodes are added, the amplitudes further along the chain appear to converge towards their original values.
Phase differences are largely unchanged along the chain, with differences of only around 5\% at the two extremes of perturbations to $c$. This can be seen in the bottom two subfigures of Figure \ref{F:changec}, along with the differences in amplitude of the nodes.

\subsubsection{Changing Parameter $b$ in only Node $4$}

\begin{figure}[h!]
\centerline{%
\includegraphics[width=0.4\textwidth]{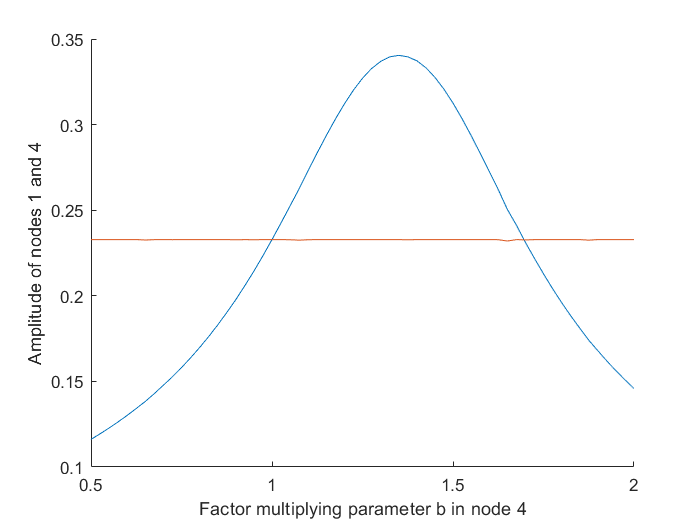} \qquad
\includegraphics[width=0.4\textwidth]{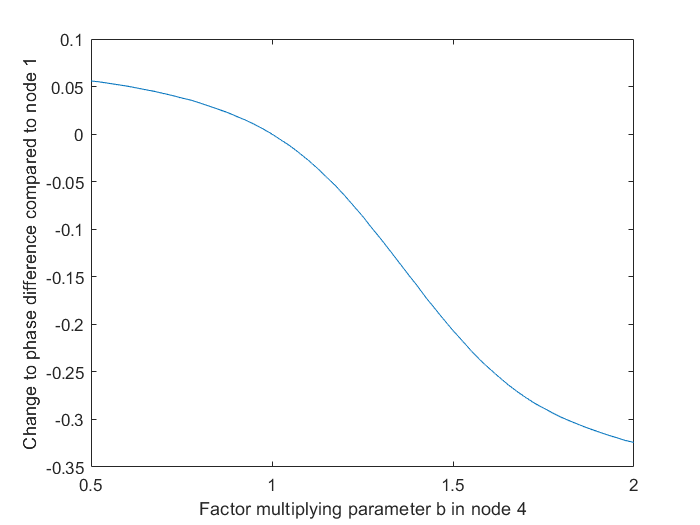}
}
\centerline{%
\includegraphics[width=0.4\textwidth]{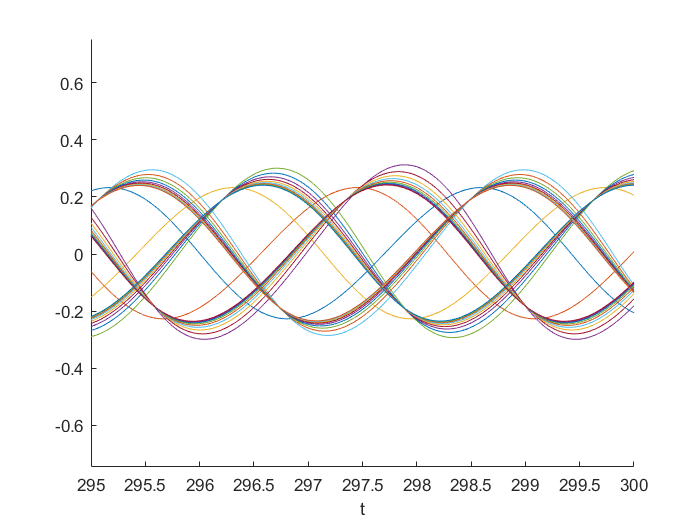}
}
\caption{Graphs showing ({\em top left}) the effect of altering the parameter $b$ in node 4
on the amplitude of node 4, and ({\em top right}) phase difference between nodes 4 and 1.
{\em Bottom}: the feedforward network extended to 
30 nodes, whose time series are
superposed, when parameter $b$ is multiplied by a factor of $1.5$.
}
\label{F:changeb}
\end{figure}

As with the case of changing the coupling, the most noticable change to patterns of oscillation when changing parameter $b$ in node 4 is to the amplitude of that node. The difference is that at the two extremes the amplitude is decreased (to 0.116 and 0.145 respectively) but as $b$ is increased this rises to a value larger than the unperturbed case before falling again, reaching a maximum of 0.337 approximately around a multiplying factor of 1.4 (Figure \ref{F:changeb}, top left).

Between a factor of 0.5 and 1.1 the phase difference between nodes 1 and 4 remains largely unchanged, but as $b$ is increased further the phase difference starts to drift. By the time $b$ is multiplied by a factor of 2 the phase difference is 0.676, and it seems to be converging to become in phase with node 2 (see Figure \ref{F:changeb}, top right). However, this apparent convergence is an artifact of stopping at a factor of 2. When $b$ is increased further, the phase difference continues to increase, albeit more slowly. A more important feature, when thinking of feedforward chains as a mechanism to propagate patterns, is that although the phase difference between node 1 and 4 is changing, the phase difference between successive nodes in the chain beyond node 4 remains fairly consistently within a few per cent of the unperturbed value $0.667$.

Figure \ref{F:changeb} (bottom) shows an extended chain of 30 nodes whose time series are  superposed. Three clumps of approximately in-phase oscillations are evident, as well as
 changes in amplitude further downstream. These clumps are separated by 
 phase differences close to one third of a period.

\subsubsection{Changing Parameter $\gamma$ in only Node $4$}

\begin{figure}[h!]
\centerline{%
\includegraphics[width=0.4\textwidth]{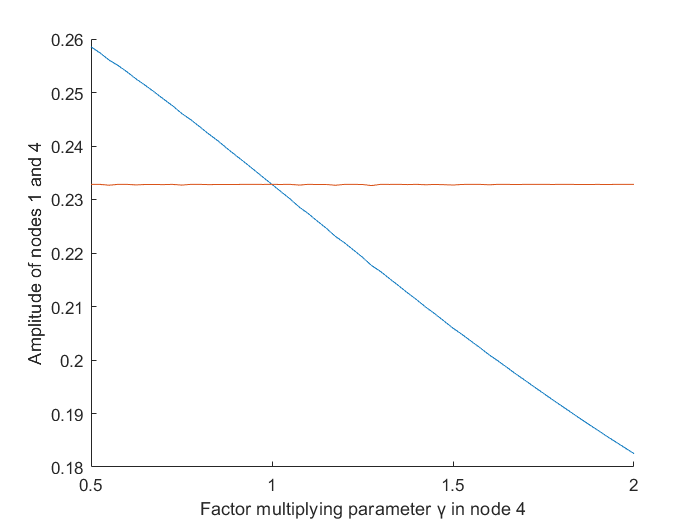}
}
\caption{Graph showing the effect on the amplitude of node 4 ({\em blue}) compared to that of nodes 1 to 3 ({\em red}) when changing the parameter $\gamma$ in node 4.
}
\label{F:changegamma}
\end{figure}

Finally, changing the value of parameter $\gamma$ in node 4, again by a factor of 0.5 to 2, also changes the maximum amplitude of node 4 (from 0.258 down to 0.183) monotonically 
as $\gamma$ increases. In contrast, changes to phase differences are minimal, within 3\% of the normal value at the two extremes (see Figure \ref{F:changegamma}). Again, the amplitudes of the upstream nodes converge back towards their unperturbed values.

\subsubsection{Other Modifications}

We chose the 3-node ring as one of the simplest GPGs that
generates periodic discrete rotating waves, leading to traveling waves in the chain.
The length of the chain, here 4, is an arbitrary choice that illustrates the
main points and is unrelated to the number of nodes in the CPG.
These choices were made as `proof of concept' rather than for their
biological significance. In further work \cite{SW23c} we apply feedforward
lifts to animal locomotion, using more complex and more realistic topologies for both
the CPG and the chain.

Experience suggests that similar results can be expected
for any CPG with cyclic group symmetry
that has periodic states with corresponding phase relations. Experiments
with rings with more than 3 nodes (not reported here)
are consistent with this expectation for a broad range of model ODEs.
The topology of the `chain' can be more complex; all that is needed is
the feedforward lift structure.

\section{Conclusions}
\label{S:C}

Propagation of synchronous or phase-synchronous states along
linear chains is important in biology, medicine, and robotics. Such signals can be generated
by a CPG and propagated along a feedforward chain.

Given the CPG, many different architectures
for propagation are possible, but those with short-range connections
are particularly useful and natural, both in engineering and evolutionary terms. 

An important issue is the stability of the propagating signals. This was discussed
in \cite{SW23a} in terms of four stability notions: transverse stability of the synchrony subspace,
and transverse Liapunov, asymptotic, or Floquet stability of the periodic orbit.
Analytic calculations and simulations show that
four standard models of neurons---namely FitzHugh--Nagumo,
Morris--Lecar, Hodgkin--Huxley, and Hind\-marsh--Rose neurons---satisfy transverse stability 
conditions, either universally or over broad parameter ranges. They can
satisfy the transverse Floquet condition even when the synchrony subspace
is transversely unstable at some points of the periodic orbit.

Simulations and some theoretical considerations suggest that 
a more general condition, transverse stability on average, might also
be a useful indicative test for stability, perhaps in a weaker sense 
such as the `Milnor attractor' condition that most nearby initial conditions 
converge to the orbit \cite{M85}. 

Real-world systems break
the idealized conditions assumed in the theoretical model.
Simulations suggest that  breaking synchrony through a variety of different modifications
leads to states that approximate the idealized 
ones in many important respects, provided the modifications are not too great.

Phase relations in periodic signals appear to be more robust to such modifications than
amplitudes.

Feedforward lifts provide a simple, effective, and robust way to propagate signals
with specific synchrony and phase patterns in a stable manner.

\end{document}